\documentclass[a4paper,fleqn,usenatbib]{mnras}

\usepackage{newtxtext,newtxmath}

\usepackage[T1]{fontenc}
\usepackage{ae,aecompl}

\usepackage{graphicx}	
\usepackage{amsmath}	
\usepackage{amssymb}	
\usepackage{color}
\usepackage{comment}
\usepackage{multirow}

\title[Detectability with 21cm-LAE cross-correlation. I\hspace{-.1em}I\hspace{-.1em}I.]{Detectability of 21cm signal during the Epoch of Reionization with 21cm-Lyman-$\alpha$ emitter\\ cross-correlation. I\hspace{-.1em}I\hspace{-.1em}I. Model dependence.}
\author[K. Kubota, A. K. Inoue, K. Hasegawa, K. Takahashi.]{
 Kenji Kubota$^1$\thanks{E-mail:175d9001@st.kumamoto-u.ac.jp},
 Akio K. Inoue$^{3,4,5}$,
 Kenji Hasegawa$^6$,
 Keitaro Takahashi$^{1,2}$
\\
$^{1}$Graduate School of Science and Technology (GSST), Kumamoto University, 2-39-1 Kurokami, Kumamoto 860-8555, Japan\\
$^{2}$International Research Organization for Advanced Science and Technology, Kumamoto University\\
$^{3}$Department of Environmental Science and Technology, Faculty of Design Technology, Osaka Sangyo University, 3-1-1\\ Nakagaito, Daito, Osaka 574-8530, Japan\\
$^{4}$Department of Physics, School of Advanced Science and Engineering, Waseda University, 3-4-1 Okubo, Shinjuku, Tokyo 169-8555, Japan\\
$^{5}$Waseda Research Institute for Science and Engineering, 3-4-1, Okubo, Shinjuku, Tokyo 169-8555, Japan\\
$^{6}$Department of Physics and Astrophysics, Nagoya University Furo-cho, Chikusa-ku, Nagoya, Aichi 464-8602, Japan\\
}

\date{Accepted XXX. Received YYY; in original form ZZZ}

\pubyear{2018}

\begin{document}
\label{firstpage}
\pagerange{\pageref{firstpage}--\pageref{lastpage}}
\maketitle

\begin{abstract}
\ Detecting $\ion{H}{i}$ 21cm line in the intergalactic medium (IGM) during the Epoch of Reionization (EoR) suffers from foreground contamination such as Galactic synchrotron and extragalactic radio sources. Cross-correlation between the 21cm line and Lyman-$\alpha$ emitter (LAE) galaxies is a powerful tool to identify the 21cm signal since the 21cm line emission has correlation with LAEs while the LAEs are statistically independent of the foregrounds. So far, the detectability of 21cm-LAE cross-power spectrum has been investigated with simple LAE models where the observed Ly$\alpha$ luminosity is proportional to the dark matter halo mass. However, the previous models were inconsistent with the latest observational data of LAEs obtained with Subaru/Hyper Suprime-Cam (HSC). Here, we revisit the detectability of 21cm-LAE cross-power spectrum adopting a state-of-the-art LAE model consistent with all Subaru/HSC observations such as the Ly$\alpha$ luminosity function, LAE angular auto-correlation, and the LAE fractions in the continuum selected galaxies. We find that resultant cross-power spectrum with the updated LAE model is reduced at small scales ($k\sim1\ \rm Mpc^{-1}$) compared to the simple models,  while the amplitudes at large scales ($k \lesssim 0.2 \ \rm Mpc^{-1}$) are not affected so much. We conclude that the large-scale signal would be detectable with Square Kilometre Array (SKA) and HSC LAE cross-correlation but detecting the small scale signal would require an extended HSC LAE survey with an area of $\sim 75\ \rm deg^2$ or 3000 hrs observation time of 21cm line with SKA.
\end{abstract}




\section{Introduction}

\ \ Understanding the evolution of the epoch of reionization (EoR) is a clue to reveal the nature and evolution of first stars and galaxies.
The EoR has been probed by the Gunn-Peterson test \citep{1965ApJ...142.1633G} in the spectra of high redshift quasars, and the integrated Thomson scattering optical depth of CMB photons.
The former indicates the EoR completed by $z\sim6$ \citep{2006AJ....132..117F} and the latter implies the reionization redshift $z\sim7.7$ if an instantaneous reionization scenario is assumed \citep{2018arXiv180706209P}.
Recently, the project called "Systematic Identification of LAEs for Visible Exploration and Reionization Research Using Subaru/HSC" (SILVERRUSH) reported a large sample of $\sim 2,000$ Lyman-$\alpha$ emitters (LAEs) at $z=5.7$ and $6.6$ (\citealt{2018PASJ...70S..13O}, \citealt{2018PASJ...70S..14S}, \citealt{2018PASJ...70S..15S}, \citealt{2018PASJ...70S..16K}, \citealt{2018ApJ...859...84H}, \citealt{2018PASJ...70...55I}, \citealt{2018arXiv180100531H}) and estimated the neutral hydrogen fraction to be $x_{\ion{H}{i}}=0.3\pm0.2$ at $z=6.6$ by comparing the Ly$\alpha$ luminosity function measurements with the observational data and reionization models.

Observing the redshifted 21cm line from the neutral IGM is the powerful way to understand the evolution of the EoR.
However, the EoR 21cm signal is much weaker than foregrounds such as Galactic synchrotron and extragalactic radio emissions.
The ongoing radio interferometers such as the Murchison Widefield Array (MWA) \citep{2009IEEEP..97.1497L,2013PASA...30....7T,2013MNRAS.429L...5B}, the LOw Frequency ARray (LOFAR) \citep{2013A&A...556A...2V,2013MNRAS.435..460J}, Hydrogen Epoch of Reionization Array (HERA) \citep{2016icea.confE...2D} and the Precision Array for Probing the Epoch of Reionization (PAPER) \citep{Jacobs2015,2015ApJ...809...61A} suffer from the foregrounds and the EoR 21cm signal has never been detected so far.
 \citet{2018Natur.555...67B} reported the first detection of the 21cm global signal during the Cosmic Dawn with the Experiment to Detect the Global EoR Signature (EDGES), but detecting the 21cm signal during the EoR is still challenging.
The Square Kilometre Array LOW (SKA-LOW) \citep{2015aska.confE.171C} will have enough sensitivity to detect the 21cm signal, but the identification of 21cm-line signal is still hard after subtracting and/or avoiding the foregrounds.

To identify the 21cm signal from the contaminated data, the cross-correlation between the 21cm line and LAEs is expected to be effective \citep{2009ApJ...690..252L,2013MNRAS.432.2615W,2014MNRAS.438.2474P,2016MNRAS.459.2741S,2016MNRAS.457..666V,2016arXiv160501734H,2016arXiv161109682H,2017arXiv170107005F,2018MNRAS.479.2754K,2018MNRAS.479.2767Y,2018MNRAS.479L.129H,2019arXiv191111783W}.
The 21cm signal has a spatial correlation with the LAEs while the foregrounds are not correlated with the LAEs.
In our previous work (\citealt{2018MNRAS.479.2754K}, Paper I), we have investigated the intrinsic detectability of the 21cm-LAE cross-power spectrum combining the 21cm observation of the MWA and SKA with the LAE survey by Subaru/HSC.
We concluded that both the MWA and SKA have an ability to detect the signal and proposed the strategies to enhance the detectability (Paper I).
Further, we studied the effects of the foregrounds which contribute to the variance, rather than the mean, of noises (\citealt{2018MNRAS.479.2767Y}, Paper II).
This paper is the third paper in a series of \citet{2018MNRAS.479.2754K} and \citet{2018MNRAS.479.2767Y}.

However, the LAE models in previous papers including our Papers I and II were rather simple in that the observed Ly$\alpha$ luminosity is proportional to the dark matter halo mass.
In reality, the Ly$\alpha$ luminosity depends on the nature of the stars and a state of the interstellar medium (ISM) \citep{2014MNRAS.441.2861H,2015MNRAS.450.4025H,2015MNRAS.453.1843S}.
A recent paper \citet{2018PASJ...70...55I} has constructed LAE models by properly considering the stochastic processes of Ly$\alpha$ production, Ly$\alpha$ escape fraction, and its dependence on the halo mass.
Their `{\it best}' LAE model can explain all Subaru/HSC survey results such as the Ly$\alpha$ luminosity function, LAE angular auto-correlation, and the LAE fractions in the continuum selected galaxies.

In this third paper of our series, we investigate the LAE model dependence of the 21cm-LAE cross-power spectrum and revisit the detectability using the above LAE model.
We assess that for the photometric LAE samples and spectroscopic LAE samples, respectively.
The HSC LAE catalogue consists of photometric LAE samples identified according to the standard color-magnitude criteria from narrow and broad band images.
Although the previous studies often adopted the photometric LAE samples, the photometric LAE sample could be contaminated by slightly lower redshift objects.
To reduce the contamination, we use the spectroscopic LAE samples identified according to the redshift and Ly$\alpha$ equivalent width of the galaxies.
They will be provided by follow-up observations of the HSC LAE survey with Prime Focus Spectrograph \citep{2014PASJ...66R...1T}.

This paper is organized as follows.
In Section 2, we give notation of the 21cm-LAE cross-power spectrum.
In Section 3, we summarize the LAE model developed in Paper I and \citet{2018PASJ...70...55I}.
In Section 4, we describe the specifications for the 21cm telescope such as the MWA and SKA, and the LAE survey by Subaru/HSC to estimate an observational error on the cross-power spectrum.
The resultant cross-power spectrum and the impact on the detectability are presented in Section 5.
Finally, we summarize our results in Section 6.

\section{21cm-LAE cross-power spectrum}

The observable quantity in 21cm observation is given by brightness temperature \citep{2006PhR...433..181F},
\begin{equation}
\delta T_b(z) \approx 27 x_{\rm \ion{H}{i}} (1+\delta_{\rm m}) \left( \frac{1+z}{10} \frac{0.15}{\Omega_{\rm m} h^2} \right)^{\frac{1}{2}} \left( \frac{\Omega_{\rm b} h^2}{0.023} \right)~[\rm mK],
\label{dTb}
\end{equation}
where $x_{\rm \ion{H}{i}}$ is the neutral hydrogen fraction and $\delta_{\rm m}$ is the matter density fluctuation.
The 21cm-LAE cross-power spectrum $P_{\rm 21,LAE}({\bf k})$ is defined as
\begin{equation}
\langle \tilde\delta_{21}({\bf k_1}) \tilde\delta_{\rm LAE}({\bf k_2}) \rangle \equiv (2\pi)^3 \delta_D({\bf k_1+k_2}) P_{\rm 21,LAE}({\bf k_1}),
\end{equation}
where $\delta_D({\bf k})$ is the Dirac delta function.
$\tilde\delta_{21}({\bf k_1})$ and $\tilde\delta_{\rm LAE}({\bf k_2})$ are fluctuations of $\delta T_b$ and LAE number density in Fourier space, respectively.
In this paper, we consider the dimensionless cross-power spectrum:
\begin{equation}
\Delta_{\rm 21,LAE}^2(k) = \frac{k^3}{2\pi^2} P_{\rm 21,LAE}(k).
\end{equation}

\section{LAE models}
In this section, we describe our LAE models based on \citet{2018PASJ...70...55I}, where a varaiety of LAE models named Model A - H were presented.
The LAE model in Paper I is essentially the same as Model A, which is the simplest among Models A - H, but different parameter values were used.
\citet{2018PASJ...70...55I} has shown that Model G can explain all observation results of SILVERRUSH such as Ly$\alpha$ luminosity function, LAE angular auto-correlation function, and LAE fraction of SILVERRUSH data.
We will compare the cross power spectra of Paper I model and Model G mainly, but also discuss Model C and Model E to see their dependence on LAE properties in detail.

We use the same reionization simulations as Paper I to model LAE distribution.
In the simulations, we solve radiative transfer of ionizing photons in N-body simulation box of $(160~{\rm Mpc})^3$ combining cosmological radiative hydrodynamics (RHD) simulation.
The RHD simulation is adapted to make recipes for the intrinsic Ly$\alpha$ luminosity, the Lyman continuum escape fraction, and the IGM clumping factor \citep{2016arXiv160301961H}.
Our reionization simulation well reproduces the observational results such as the IGM neutral fraction at $z=6.6$ and CMB Thomson optical depth.
Similar to Paper I, we perform two reionization simulations named the `mid' model and `late' model.
These models have different ionizing photon production rates, and the rate of the `late' model is 1.5 times lower than that of the `mid' model so that the completion of reionization is delayed.
More details for the simulation are described in Paper I and will be provided by Hasegawa et al. (in preparation).
The mock LAE samples are generated from the N-body halos using the RHD recipes via two steps.
Firstly, we compute intrinsic Ly$\alpha$ luminosity of each galaxy.
Secondly, we estimate observable Ly$\alpha$ luminosity of each galaxy by considering the escape fraction of Ly$\alpha$ photons and attenuation of Ly$\alpha$ photons through the IGM.

\subsection{Paper I model}

In Paper I model, the intrinsic Ly$\alpha$ luminosity is computed by using one-to-one relation between the intrinsic Ly$\alpha$ luminosity and the halo mass,
\begin{equation}
	L^{\rm int}_{\alpha,42}= M_{\rm h,10}^{1.1},
\label{L_int}
\end{equation}
where $L^{\rm int}_{\alpha,42}$ is the intrinsic Ly$\alpha$ luminosity normalized with $10^{42}\ \rm erg/s$ and $M_{\rm h,10}$ is a halo mass normalized with $10^{10}\ \rm M_{\rm \odot}$. Then, we estimate the observable Ly$\alpha$ luminosity of each galaxy with,
\begin{equation}
	L^{\rm obs}_{\alpha,42} = f_{\rm esc, \alpha}T_{\alpha, \rm IGM}L^{\rm int}_{\alpha,42},
\label{L_obs}
\end{equation}
where $f_{\rm esc, \alpha}$ and $T_{\alpha, \rm IGM}$ are the escape fraction of Ly$\alpha$ photons and transmission of Ly$\alpha$ photons through the IGM, respectively.
$f_{\rm esc, \alpha}$ is a model parameter and it is set to be consistent with the Ly$\alpha$ luminosity function of \citet{2014ApJ...797...16K} and \citet{2018PASJ...70S..16K}.
Consequently, it is set as $f_{\rm esc, \alpha}=0.25\ (0.40)$ in the `mid (late)' model.
$T_{\alpha, \rm IGM}$ is sensitive to a line profile $\phi_{\alpha}(\nu)$ emerging from the surface of a galaxy.
To calculate $T_{\alpha, \rm IGM}$, we determine the line profile from Ly$\alpha$ radiative transfer with an expanding spherical cloud model \citep{2018MNRAS.477.5406Y}.
We have assumed $150\ {\rm km\ s^{-1}}$ and $10^{19}\ {\rm cm^{-2}}$ for the velocity and the column density, respectively.
The line profile depends on the galactic wind velocity and the $\ion{H}{i}$ column density in a galaxy.
In the expanding cloud model, Ly$\alpha$ photons with shorter wavelengths are selectively scattered by outflowing gas.
It results in an asymmetric line profile with a characteristic peak at $\lambda>1216\ \AA$.
Then, $T_{\alpha, \rm IGM}$ is calculated as,
\begin{equation}
	T_{\alpha, \rm IGM} = 
	\frac{\int \phi_{\alpha}(\nu _0)~e^{-\tau_{{\nu_0},\rm IGM}} d\nu_0}
	{\int \phi_{\alpha}(\nu_0) d\nu_0}\,,
\end{equation}
where $\nu_0$ is the frequency in the rest-frame of a galaxy and $\tau_{\nu,\rm IGM}$ is the optical depth of Ly$\alpha$ photons through the IGM.
$\tau_{\nu,\rm IGM}$ is computed by integrating the Ly$\alpha$ cross section $s_{\alpha}$ of neutral hydrogen with respect to the distance from an LAE candidate in the physical coordinate,
\begin{equation}
	\tau_{\nu_0, \rm IGM} = \int_{r_{\rm vir}}^{l_{\rm p,max}} 
	s_\alpha(\nu,T_{\rm g}) n_{\rm \ion{H}{i}} dl_{\rm p}.
\end{equation}
The integration is performed from the virial radius of the halo ($r_{\rm vir}$) to the maximum distance of $l_{\rm p,max} = 80\ \rm cMpc$, and the choice of the maximum distance has a negligible effect if we take a large enough value.
Once the observable Ly$\alpha$ luminosity is estimated, LAEs detectable with the HSC are selected to make LAE samples (e.g. $L_{\alpha, \rm obs}\ \geq\ 4.1\times10^{42}\ \rm erg\ s^{-1}$ for HSC Deep survey).
These LAE samples correspond to the photometric samples.

\subsection{Model G}

In Paper I, we assumed that the intrinsic Ly$\alpha$ luminosity is uniquely determined by the halo mass and that the Ly$\alpha$ escape fraction is constant.
In fact, \citet{2014MNRAS.440..776Y} showed that Ly$\alpha$ escape fraction has a large dispersion due to the interaction in the ISM.
Then, in Model G of \citet{2018PASJ...70...55I}, we introduced the stochasticity in the Ly$\alpha$ photon production and transmission in galaxy halos into our LAE models.
Further, Model G has considered the halo-mass dependence in the transmission of Ly$\alpha$ photons.
These two effects, the stochasticity and halo-mass dependence concerning the production and transmission of Ly$\alpha$ photons, are new ingredients compared with Paper I model.
We summarize the recipe for Model G LAE samples below.

Firstly, to consider the stochasticity of the Ly$\alpha$ photon production, we use the following formula,
\begin{equation}
L^{\rm int}_{\alpha,42}
= (M_{\rm h,10})^{1.1} \times 10^{\delta_{\rm L_\alpha}} \times (1-e^{-10M_{\rm h,10}}),
\label{L_int_sto}
\end{equation}
instead of Eq. (\ref{L_int}).
The differences between Eqs. (\ref{L_int}) and (\ref{L_int_sto}) are the presence of the factor $10^{\delta_{\rm L_\alpha}}$ and the exponential term.
The former represents the stochastic part of the Ly$\alpha$ photon production and the value of $\delta_{\rm L_\alpha}$ is given according to a Gaussian probability distribution with the mean of zero and the standard deviation $\sigma_{L_\alpha} = 0.6-0.3 \log_{10}M_{\rm h,10}$ if $\log_{10}M_{\rm h,10} \leq 2$ and otherwise $\sigma_{L_\alpha} = 0$.
On the other hand, the exponential term explains the reduction of the Ly$\alpha$-photon production due to a high escape fraction of ionizing photons in low mass galaxies.
This is because when more ionizing photons escape into the IGM without absorption within galaxies, cascades of recombined hydrogen atoms, which produce Ly$\alpha$ photons, is suppressed (Hasegawa et al. in prep).
In fact, this effect is negligible (less than $1\%$) for the observed LAEs which are more massive than $10^{10}\ \rm M_{\odot}$.

Secondly, to consider the stochasticity of the Ly$\alpha$ transmission, we assume a Poisson process for the interaction between $\ion{H}{i}$ gas and Ly$\alpha$ photons. Then, Ly$\alpha$ photon optical depth ($\tau_{\alpha}$) in a halo follows a Gaussian probability distribution with the standard deviation equal to the mean, $\langle \tau_{\alpha}\rangle$,
\begin{equation}
P(\tau_{\alpha}) =
\frac{\exp[-(\tau_{\alpha}-\langle \tau_{\alpha}\rangle)^2/2\langle \tau_{\alpha}\rangle]}{\sqrt{2\pi\langle \tau_{\alpha}\rangle}}.
\label{p_tau}
\end{equation}
Here, we assume $\langle \tau_{\alpha} \rangle$ depends on the halo mass as,
\begin{equation}
\langle \tau_{\alpha}\rangle=\tau_{\alpha,10}{M_{\rm h,10}}^p,
\label{tau_halomass}
\end{equation}
where $\tau_{\alpha,10}$ is a model parameter and calibrated to reproduce the observed Ly$\alpha$ luminosity function at $z=5.7$ in \citet{2018PASJ...70S..16K}.
Finally, the escape fraction of Ly$\alpha$ photons is obtained by,
\begin{equation}
	f_{\rm esc, \alpha}=e^{-\tau_{\alpha}}.
\end{equation}
In \citet{2018PASJ...70...55I}, two cases, $p=0$ and $p=1/3$, are considered for the halo mass dependence.
The $p=0$ case represents no dependence and the $p=1/3$ case means that $\langle \tau_{\alpha}\rangle$ is proportional to a column density, $M_{\rm h}/R^2_{\rm vir}$.
Actually, the latter dependence is found by simulation results \citep{2014MNRAS.440..776Y}.
Finally, the observable Ly$\alpha$ luminosity is estimated by Eq.(\ref{L_obs}) to make the photometric LAE samples.

\subsection{Models C and E}

In addition to Paper I model and Model G, we consider Models C and E to assess the model dependence of the cross-correlation more in detail, although they do not explain all the LAE properties extracted from SILVERRUSH data.
Model C has considered the stochasticity in the Ly$\alpha$ escape (Eq. \ref{p_tau}) but not the halo dependence ($p=0$ in Eq.~\ref{tau_halomass}).
In this model, the Ly$\alpha$ photons in a less massive halo gain a more chance to escape into the IGM while the Ly$\alpha$ photons from a massive halo lose the chance.
\citet{2018PASJ...70...55I} have found that this effect yields a smaller amplitude of the LAE angular auto-correlation at a smaller angular separation.
Then, Model C could well reproduce SILVERRUSH data except for the LAE fraction.

Model E has introduced only the halo mass dependence of the mean Ly$\alpha$ optical depth ($p=1/3$ in Eq.~(\ref{tau_halomass})).
This model failed to reproduce the observed Ly$\alpha$ luminosity function at the bright-end due to a very high Ly$\alpha$ optical depth in a massive halo.
Model E could marginally agree with the observed LAE angular auto-correlation, but be inconsistent with the other quantities.

\subsection{Mock LAE catalog}

To construct mock LAE catalogs, we follow the prescription given in \citet{2018PASJ...70...55I}.
First, the source rest-frame equivalent width (EW) of the Ly$\alpha$ line is obtained by
\begin{equation}
EW_0
= \frac{f_{\rm esc, \alpha} T_{\alpha, \rm IGM} L_{\alpha}^{\rm int}}{L_{\lambda_\alpha}^{\rm con}}
= \frac{f_{\rm esc, \alpha} T_{\alpha, \rm IGM} L_{\alpha}^{\rm int}}{L_{\lambda_{\rm UV}}^{\rm con} \Bigl(\lambda_\alpha/\lambda_{\rm UV}\Bigr)^{\beta}}.
\end{equation}
$L_{\lambda_\alpha}^{\rm con}$ and $L_{\lambda_{\rm UV}}^{\rm con}$ are the continuum flux densities at $\lambda_\alpha=1216\ \AA$ and $\lambda_{\rm UV}\approx1500\ \AA$, respectively.
The index $\beta$ is the UV spectral slope.
The UV luminosity $M_{\rm UV}$ is simply related to the halo mass to be consistent with \citet{2014MNRAS.440..731S} simulations:
\begin{equation}
M_{\rm UV}=-17.2-2.5\log_{10}(M_{\rm h,10})+\delta_{\rm UV},
\end{equation}
where $\delta_{\rm UV}$ represents a fluctuation in the UV magnitude.
Again, a Gaussian random number is adopted for $\delta_{\rm UV}$, where the mean is zero and the standard deviation is $\sigma_{\rm UV}=0.4-0.2\log_{10}M_{\rm h,10}$ if $\log_{10}M_{\rm h,10}\leq2$, else $\sigma_{\rm UV}=0$.
For the index $\beta$, an empirical relation is adopted \citep{2014ApJ...793..115B},
\begin{equation}
\beta = - 2.05 - 0.20 (M_{\rm UV} + 19.5) + \delta_{\beta},
\end{equation}
where $\delta_{\beta}$ represents a fluctuation in $\beta$ which follows a Gaussian probability distribution with the mean of zero and the standard deviation of $\sigma_{\beta}=0.1$ \citep{2014ApJ...793..115B,2014MNRAS.440..731S}.

In the LAE survey by the HSC, photometric LAE samples are provided as the first step.
Identifying the LAEs according to the standard color-magnitude criteria from observed images is a way to make an LAE catalogue effectively.
\citet{2018PASJ...70...55I} estimated observed magnitudes of the halos through the HSC broadband and narrowband filters to generate a mock photometric catalogue, and select LAEs by the same color-magnitude criteria as the HSC survey from the mock catalogue.
However, the photometric LAE samples could be contaminated by slightly lower redshift objects which produce strong continuum spectra which mimic Ly$\alpha$ emission line.
Such contamination can exist in the observational LAE samples if they are not confirmed spectroscopically.
To avoid the contamination as much as possible, we consider spectroscopic LAE samples.
The spectroscopic observation by the PFS will provide spectroscopically confirmed LAEs from the photometrically identified LAEs.
In this paper, we select the galaxies with ${\rm EW}_0 \geq 20~\AA$ within the redshift range of $z = 6.6 \pm 0.1$ as spectroscopic samples of LAEs.
Then, we assess the 21cm-LAE cross-power spectrum for the two kinds of the LAE samples, (1) photometric samples and (2) spectroscopic samples.
The former corresponds to the case where the PFS redshift is unavailable and the latter corresponds to the case where the PFS redshift is available, respectively.

\citet{2018PASJ...70...55I} reported that Model G can explain all observational quantities such as Ly$\alpha$ luminosity function, LAE angular auto-correlation function, and LAE fraction of SILVERRUSH data. Here, it considers the dispersion of Ly$\alpha$ transmission in the halo and the halo mass dependence of that ($p=1/3$).
On the other hand, 
Model A which is equivalent to Paper I model (not considered any stochastic processes and the halo mass dependence of Ly$\alpha$ transmission) marginally explains the Ly$\alpha$ luminosity function and the LAE angular correlation function but not the LAE fraction.
Therefore, we demonstrate the LAE model dependence of the 21cm-LAE cross-power spectrum and evaluate the impact on the detectability.

\begin{figure*}
\begin{center}
    \begin{tabular}{r}
      \begin{minipage}{0.35\hsize}
        \begin{center}
          \includegraphics[width=6.5cm]{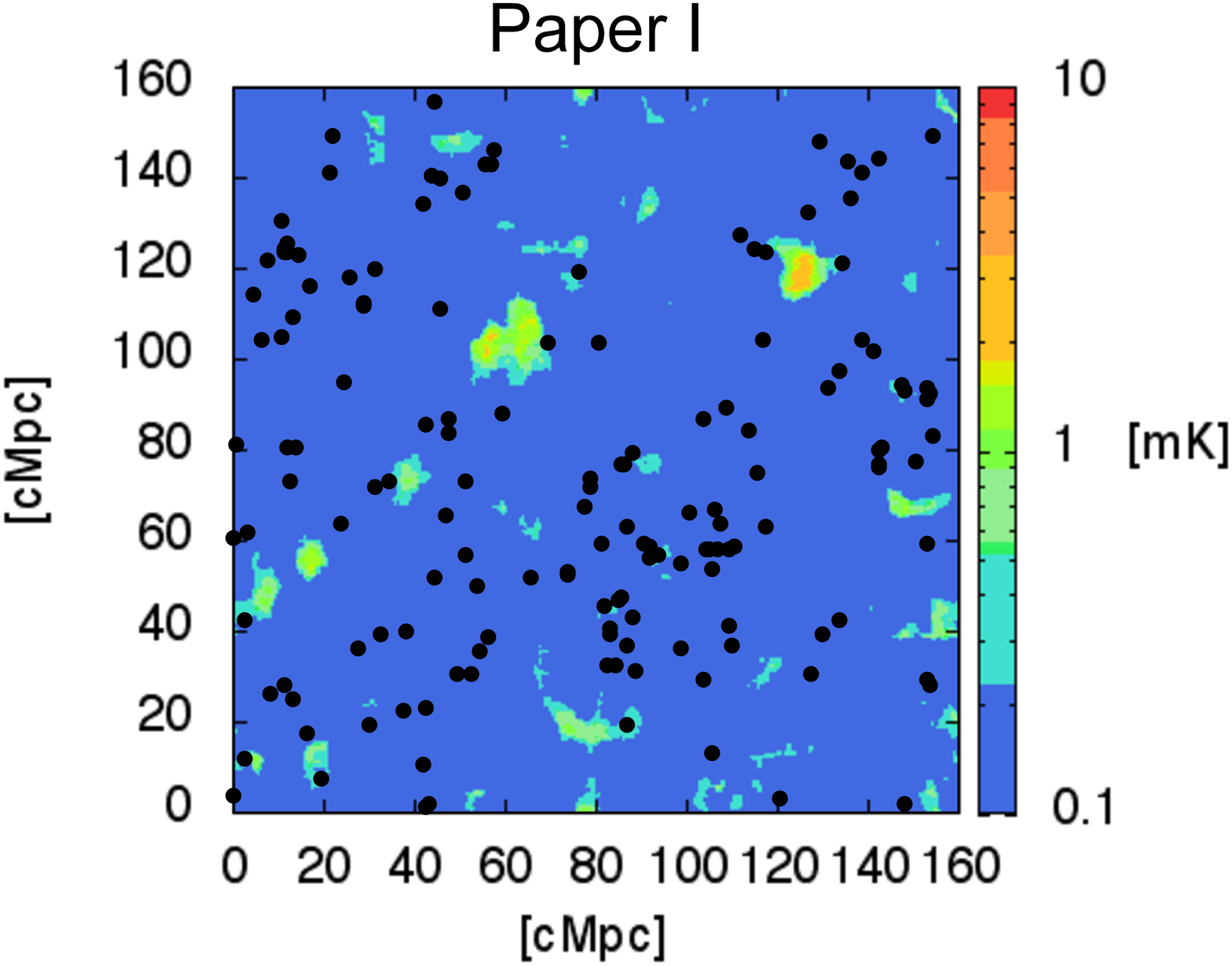}
        \end{center}
      \end{minipage}
      \begin{minipage}{0.35\hsize}
        \begin{center}
          \includegraphics[width=6.5cm]{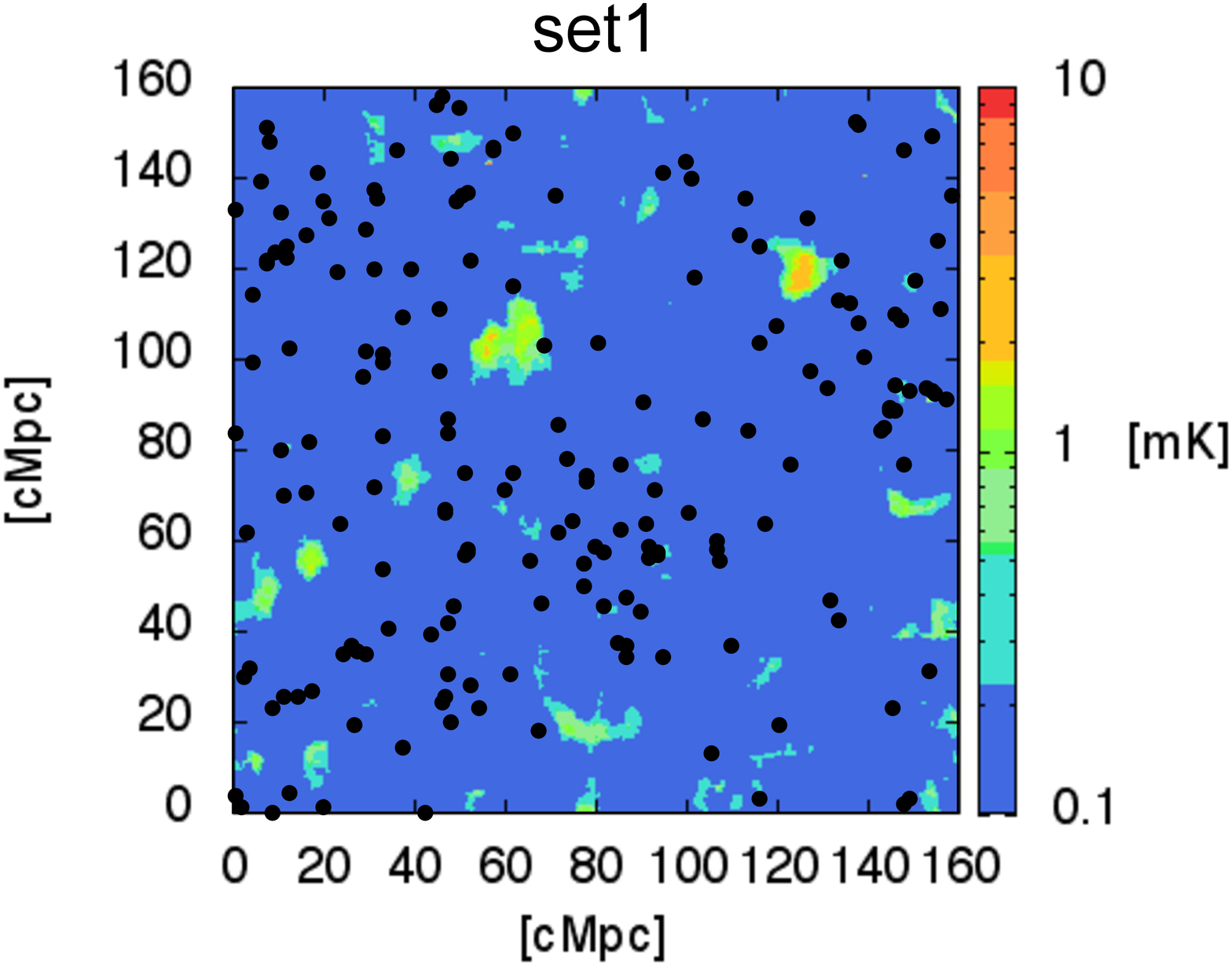}
        \end{center}
      \end{minipage}
      \begin{minipage}{0.35\hsize}
        \begin{center}
          \includegraphics[width=6.5cm]{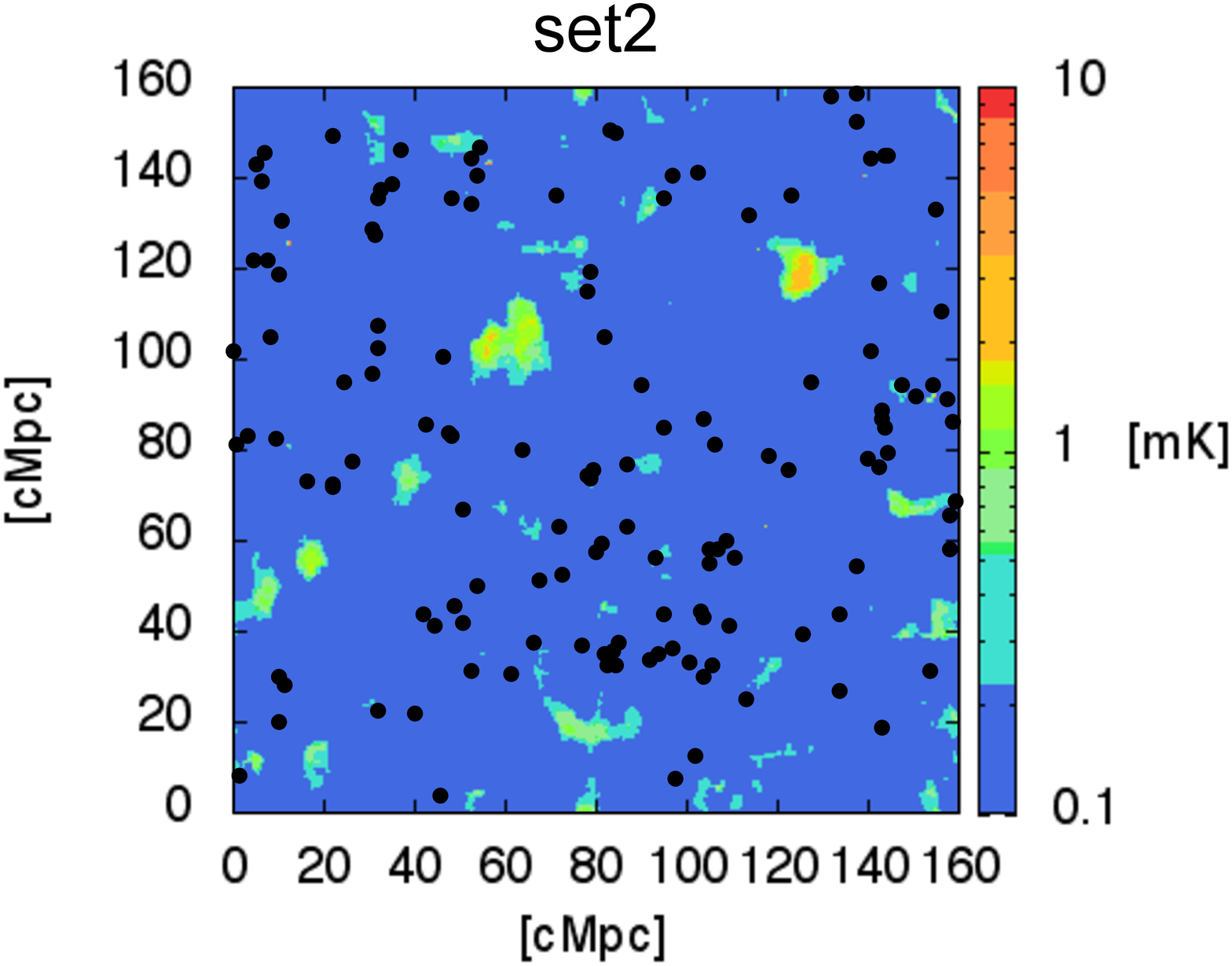}
        \end{center}
      \end{minipage}
\\      
      \begin{minipage}{0.35\hsize}
        \begin{center}
          \includegraphics[width=6.5cm]{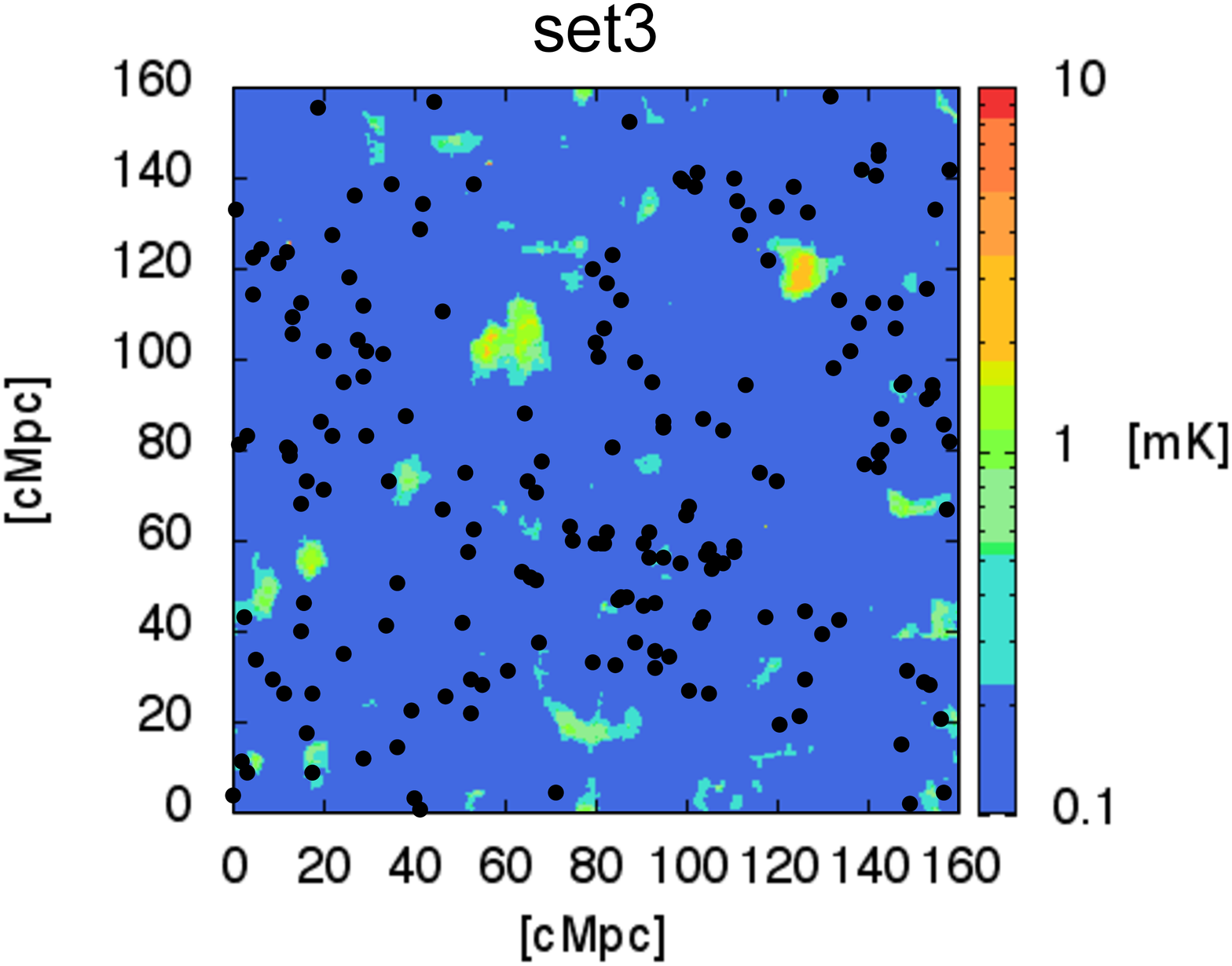}
        \end{center}
      \end{minipage}
      \begin{minipage}{0.35\hsize}
        \begin{center}
          \includegraphics[width=6.5cm]{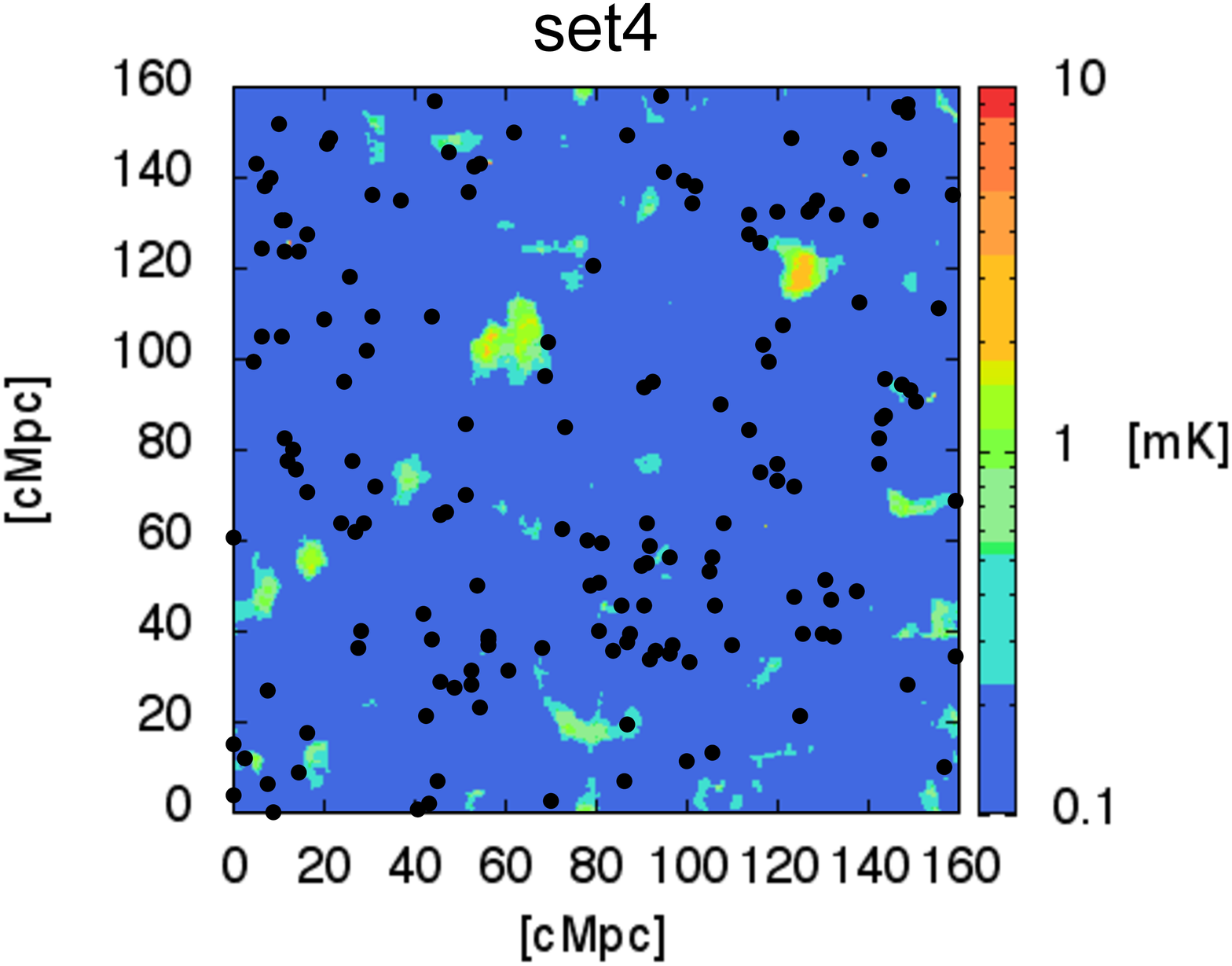}
        \end{center}
      \end{minipage}
   \end{tabular}
\caption{LAE distribution (dots) and the 21cm brightness temperature (color) for the `mid' model at $z=6.6$. The left panel shows the LAE distribution of Paper I and the right 4 panels show 4 realizations of Model G. Here, LAEs are identified by the color and magnitude of the galaxies, corresponding to observational process of photometric LAE samples. These panels are integrated with respect to the frequency within $z = 6.6 \pm 0.1$.}
\label{fig:mid_mag}
\end{center}
\end{figure*}

\begin{figure*}
\begin{center}
    \begin{tabular}{r}
      \begin{minipage}{0.35\hsize}
        \begin{center}
          \includegraphics[width=6.5cm]{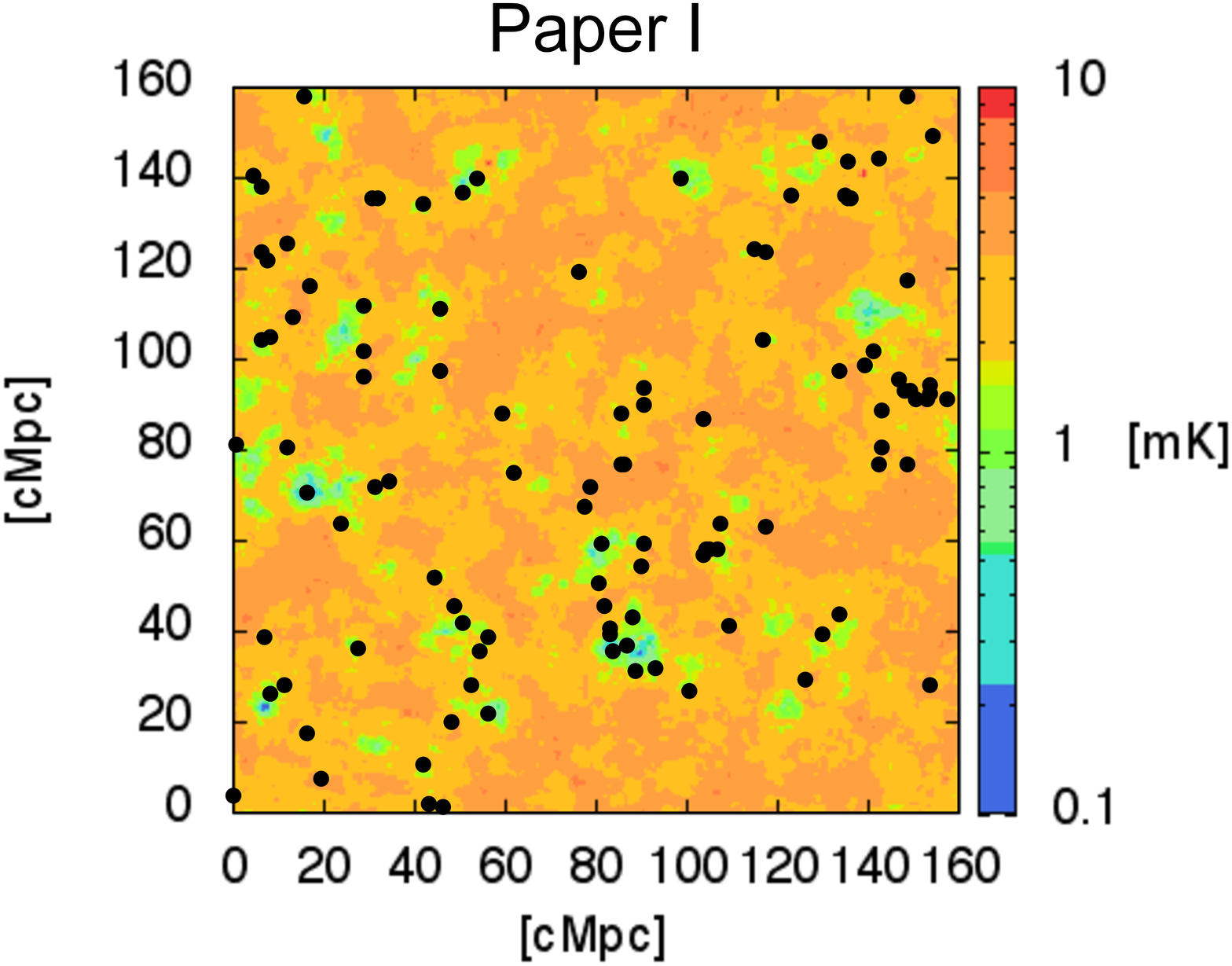}
        \end{center}
      \end{minipage}
      \begin{minipage}{0.35\hsize}
        \begin{center}
          \includegraphics[width=6.5cm]{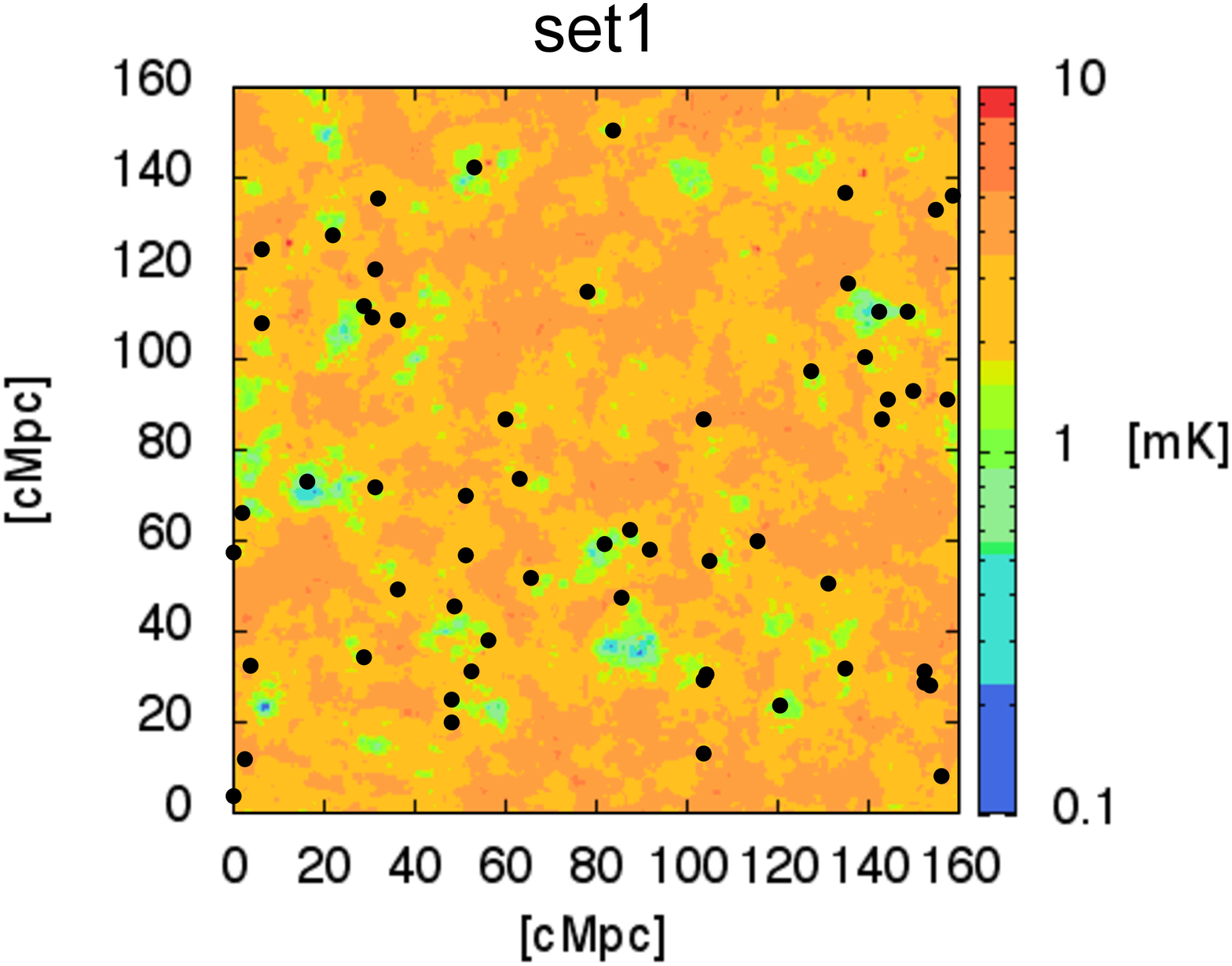}
        \end{center}
      \end{minipage}
      \begin{minipage}{0.35\hsize}
        \begin{center}
          \includegraphics[width=6.5cm]{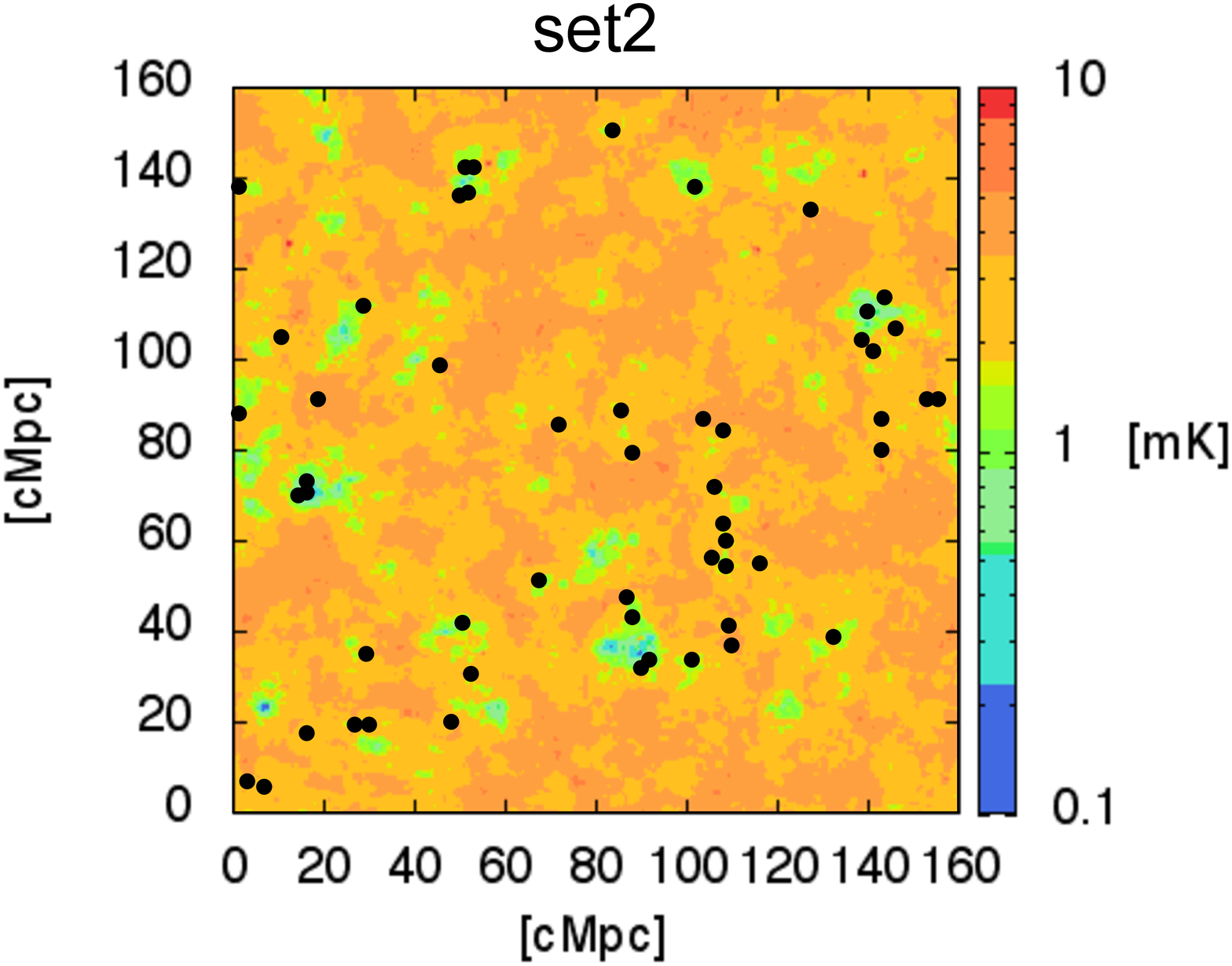}
        \end{center}
      \end{minipage}
\\      
      \begin{minipage}{0.35\hsize}
        \begin{center}
          \includegraphics[width=6.5cm]{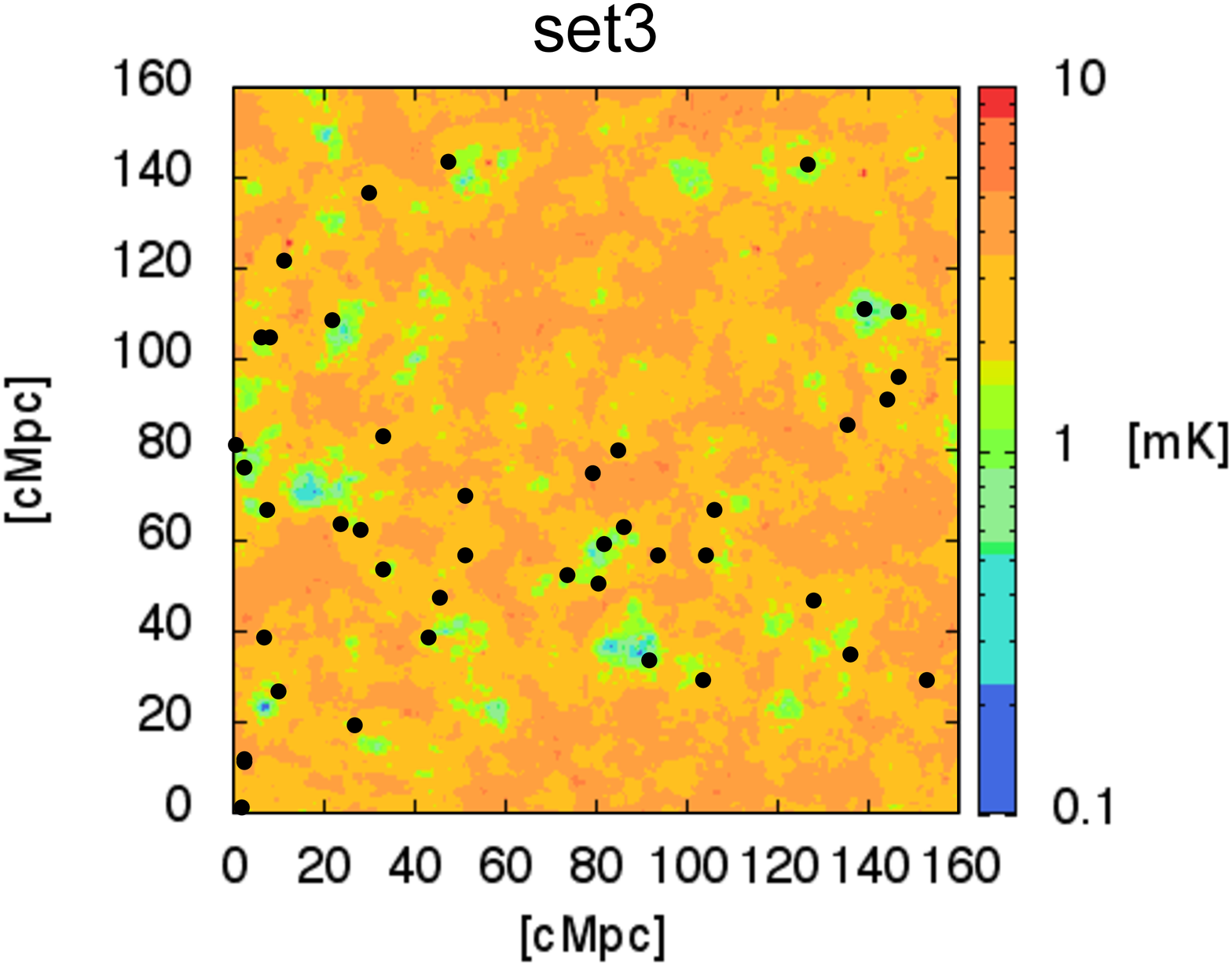}
        \end{center}
      \end{minipage}
      \begin{minipage}{0.35\hsize}
        \begin{center}
          \includegraphics[width=6.5cm]{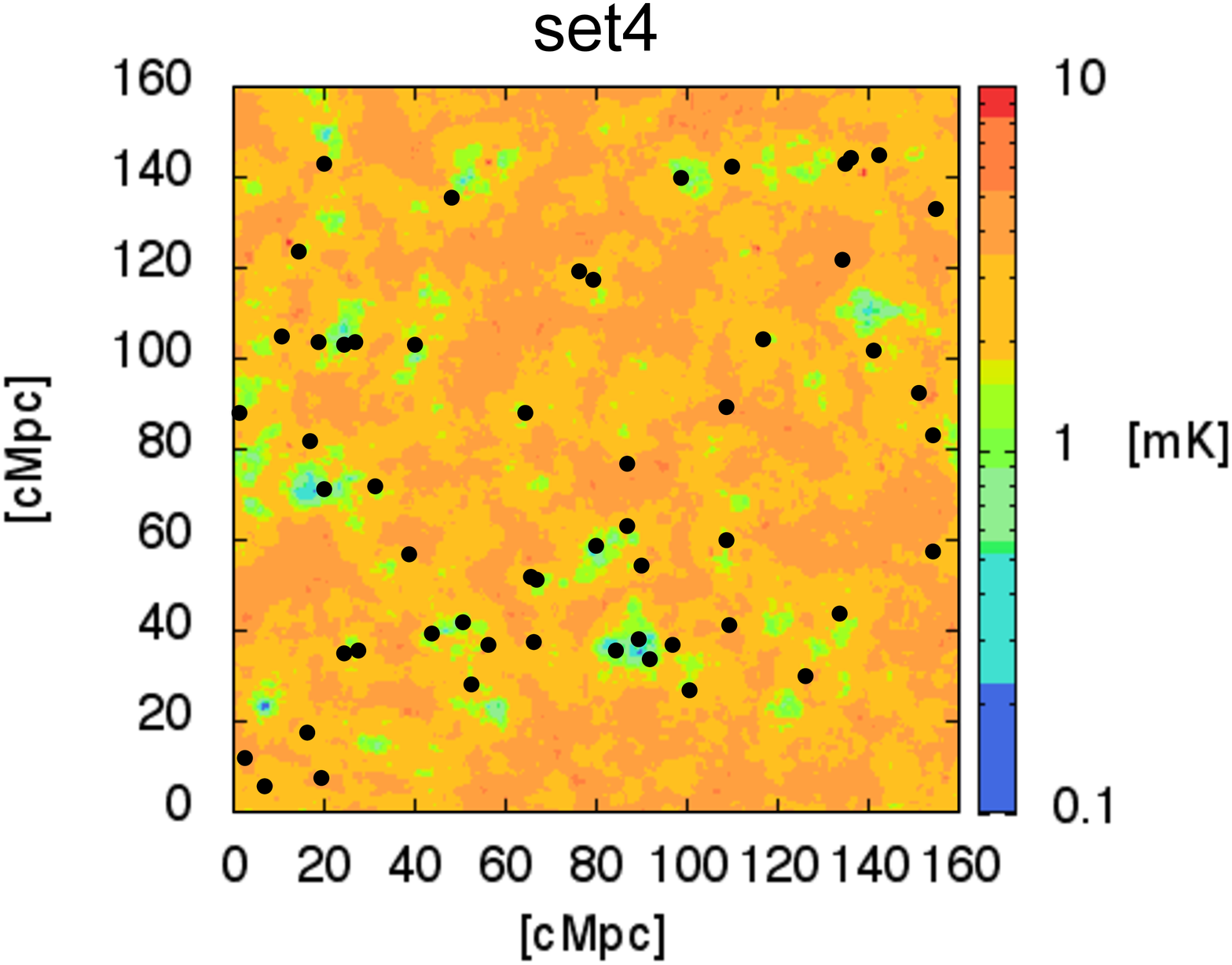}
        \end{center}
      \end{minipage}
   \end{tabular}
\caption{Same as Fig.~\ref{fig:mid_mag}, but for the `late' model.}
\label{fig:late_mag}
\end{center}
\end{figure*}

Figs.~\ref{fig:mid_mag} and \ref{fig:late_mag} show comparison of the distribution of photometric LAE samples between Paper I model and Model G for the `mid' and `late' models, respectively.
In both of the LAE models, LAEs are distributed in the ionized regions, where $\delta T_b \sim 0\ \rm mK$.
Thus, an anti-correlation between the LAE distribution and $\delta T_b$ is expected.

For the `mid' model, parameters of both Paper I model and Model G are tuned to reproduce the observed Ly$\alpha$ luminosity function.
the number of LAEs in Paper I model is consistent with Model G since the simulated Ly$\alpha$ luminosity functions reproduce the observed Ly$\alpha$ luminosity functions.
However, in the `late' model, Paper I model relatively produces larger numbers of LAEs.
This is because Paper I used different values of the Ly$\alpha$ escape fraction for the `mid' and `late' models to reproduce the observed Ly$\alpha$ luminosity function at $z=6.6$ in the both cases (see also Sec.3.1). On the other hand, \citet{2018PASJ...70...55I} chose the parameters to reproduce the observed LAE luminosity functions at $z=5.7$ in the both cases because the IGM is fully ionized at the redshift. Then, they used the same parameters at $z=6.6$ in the both cases by assuming the parameter is independent of redshift evolution and reionization models.
In fact, Model G can reproduce the observed Ly$\alpha$ luminosity function at $z=6.6$ in the 'mid' model but the simulated Ly$\alpha$ luminosity function shows a lower amplitude than the observed luminosity function in the 'late' model.
Therefore, Paper I model produces more LAEs than Model G in the 'late' model.

\begin{figure*}
\begin{center}
    \begin{tabular}{r}
      \begin{minipage}{0.35\hsize}
        \begin{center}
           \includegraphics[width=6.5cm]{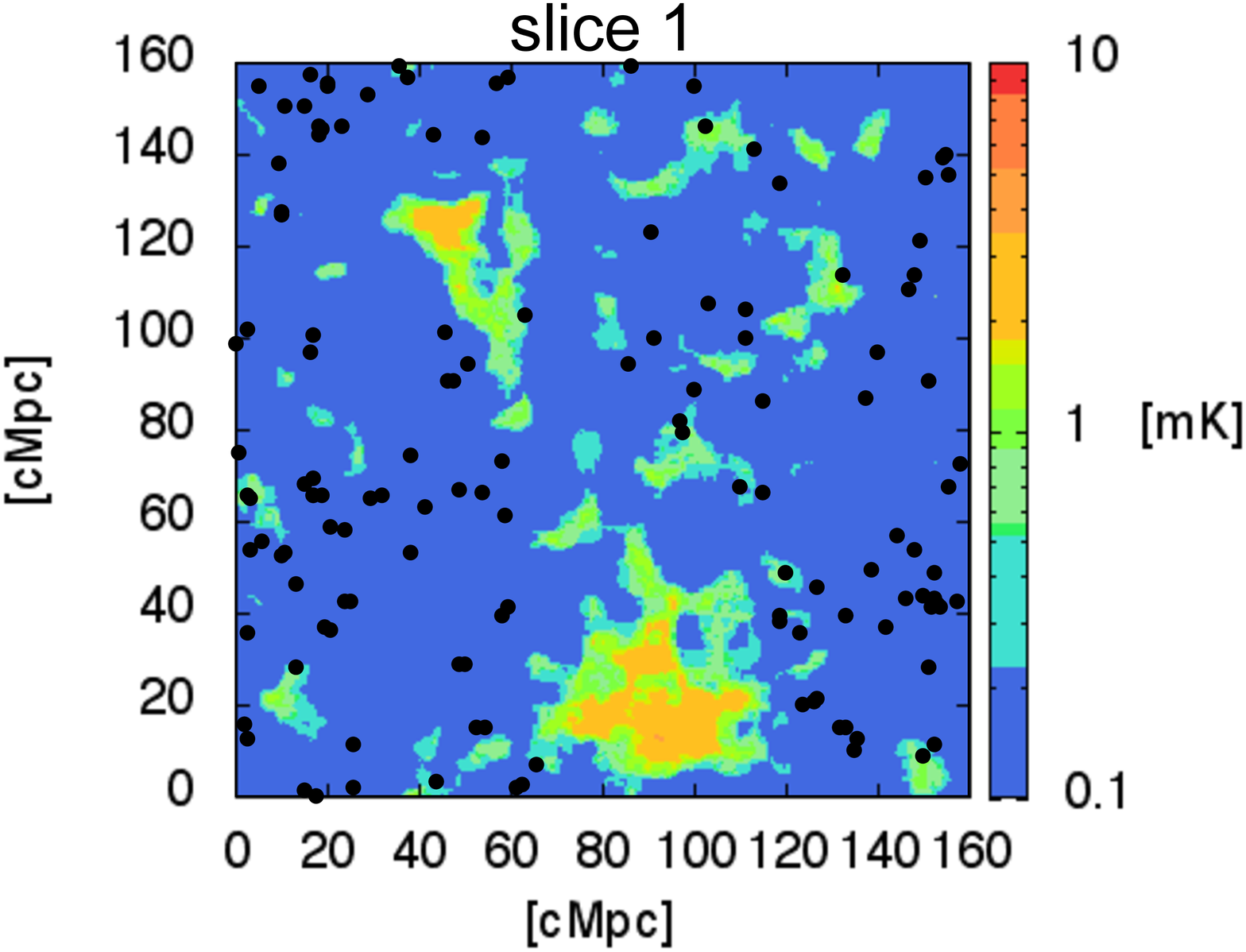}
        \end{center}
      \end{minipage}
      \begin{minipage}{0.35\hsize}
        \begin{center}
          \includegraphics[width=6.5cm]{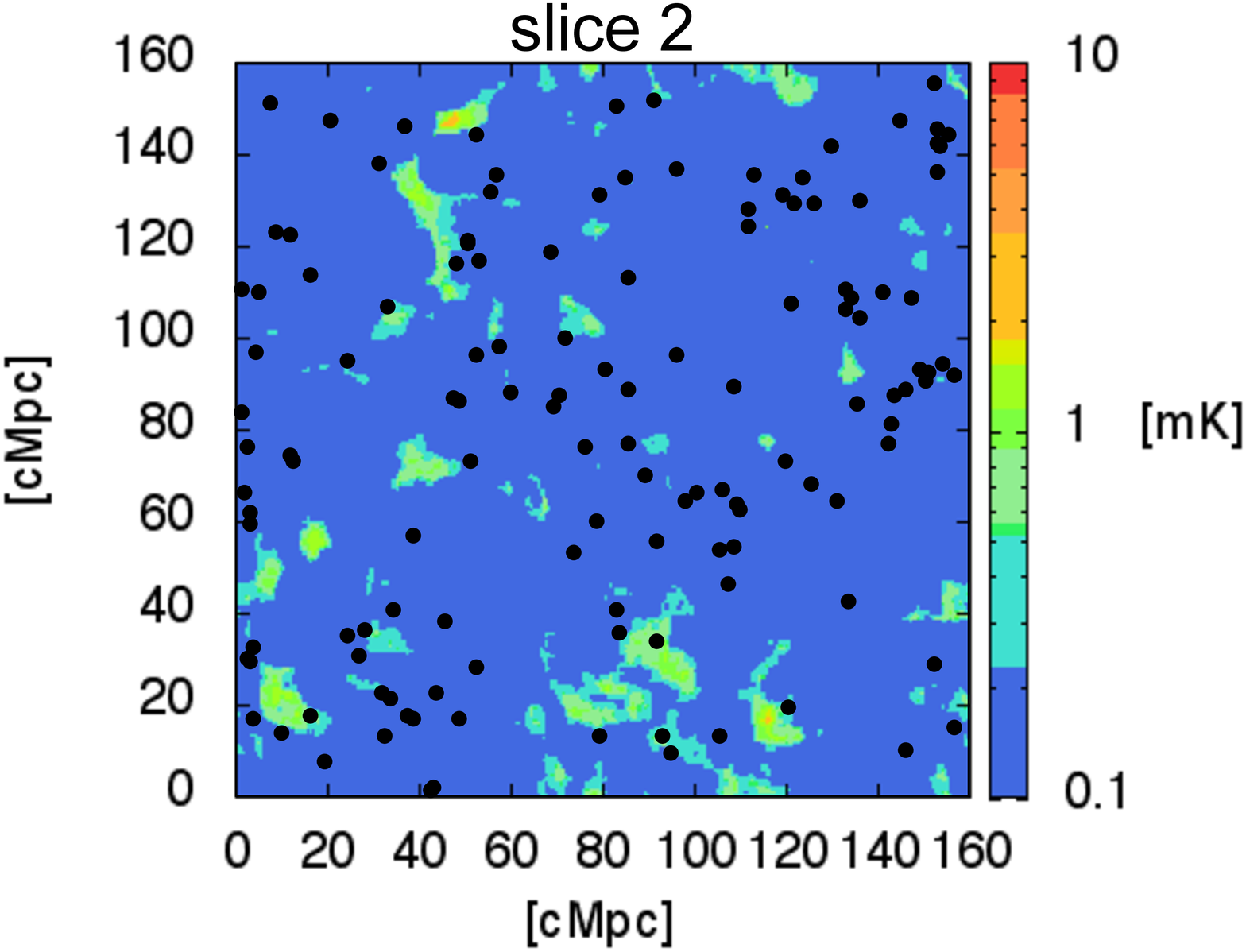}
        \end{center}
      \end{minipage}
\\      
      \begin{minipage}{0.35\hsize}
        \begin{center}
          \includegraphics[width=6.5cm]{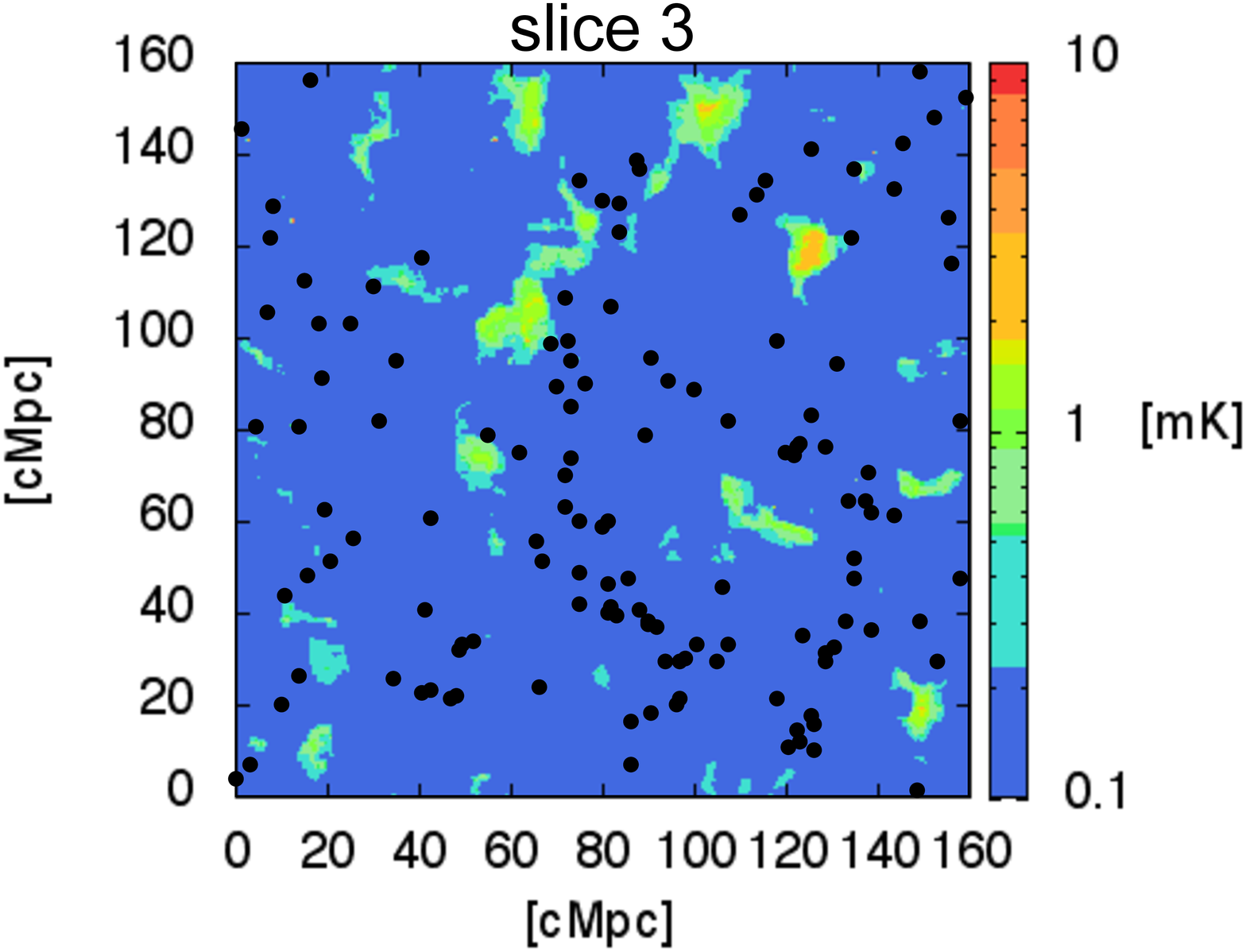}
        \end{center}
      \end{minipage}
      \begin{minipage}{0.35\hsize}
        \begin{center}
           \includegraphics[width=6.5cm]{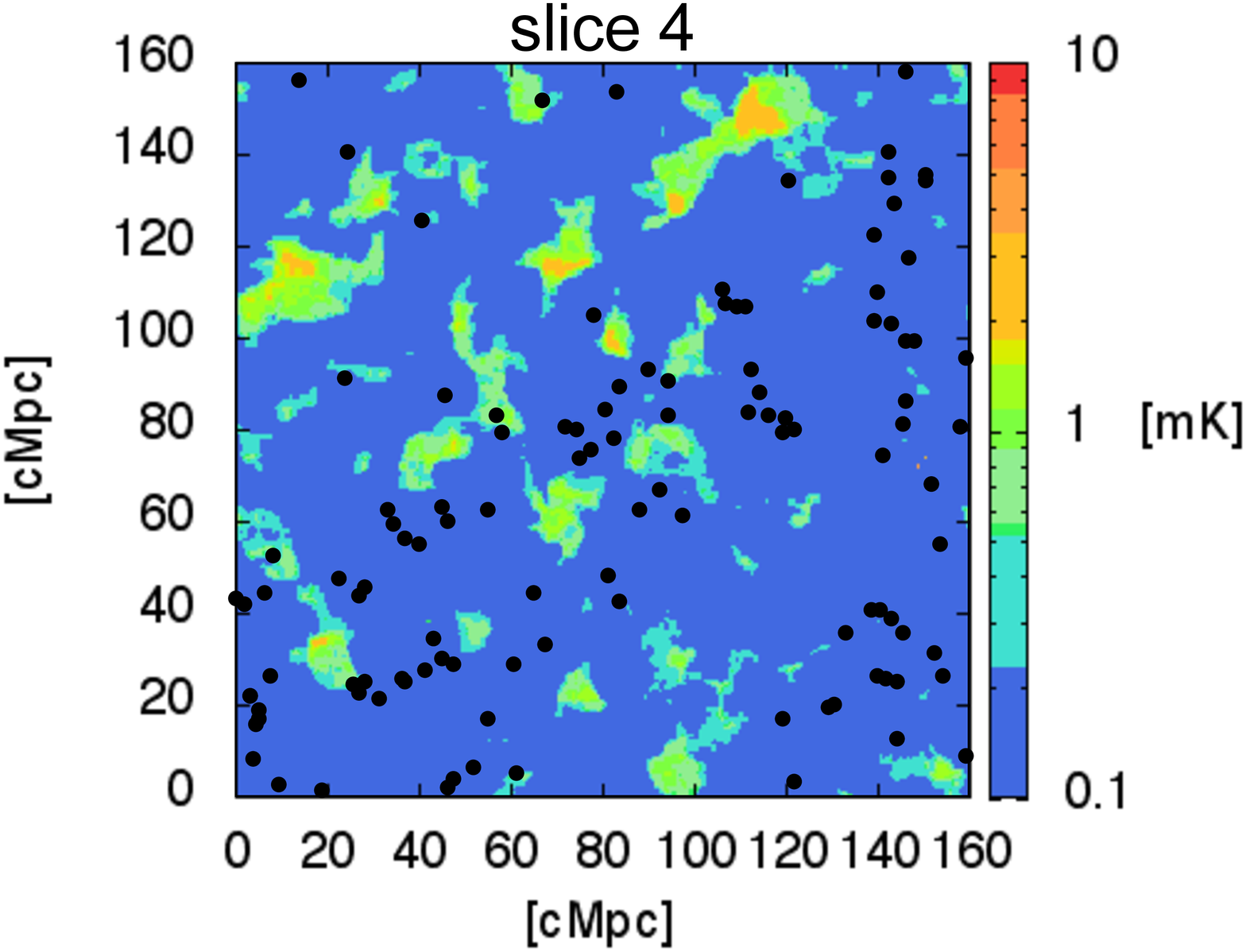}
        \end{center}
      \end{minipage}
   \end{tabular}
\caption{Same as Fig.~\ref{fig:mid_mag}, but LAEs are identified by the redshift and EW of the galaxies, corresponding to spectroscopic LAE samples observationally. Here, we demonstrate the LAE distributions of Model G (set1) in the simulation box divided into 4 blocks along $z$-axis at redshift $z=6.6$.}
\label{fig:mid_zp}
\end{center}
\end{figure*}

\begin{figure*}
\begin{center}
    \begin{tabular}{r}
      \begin{minipage}{0.35\hsize}
        \begin{center}
         \includegraphics[width=6.5cm]{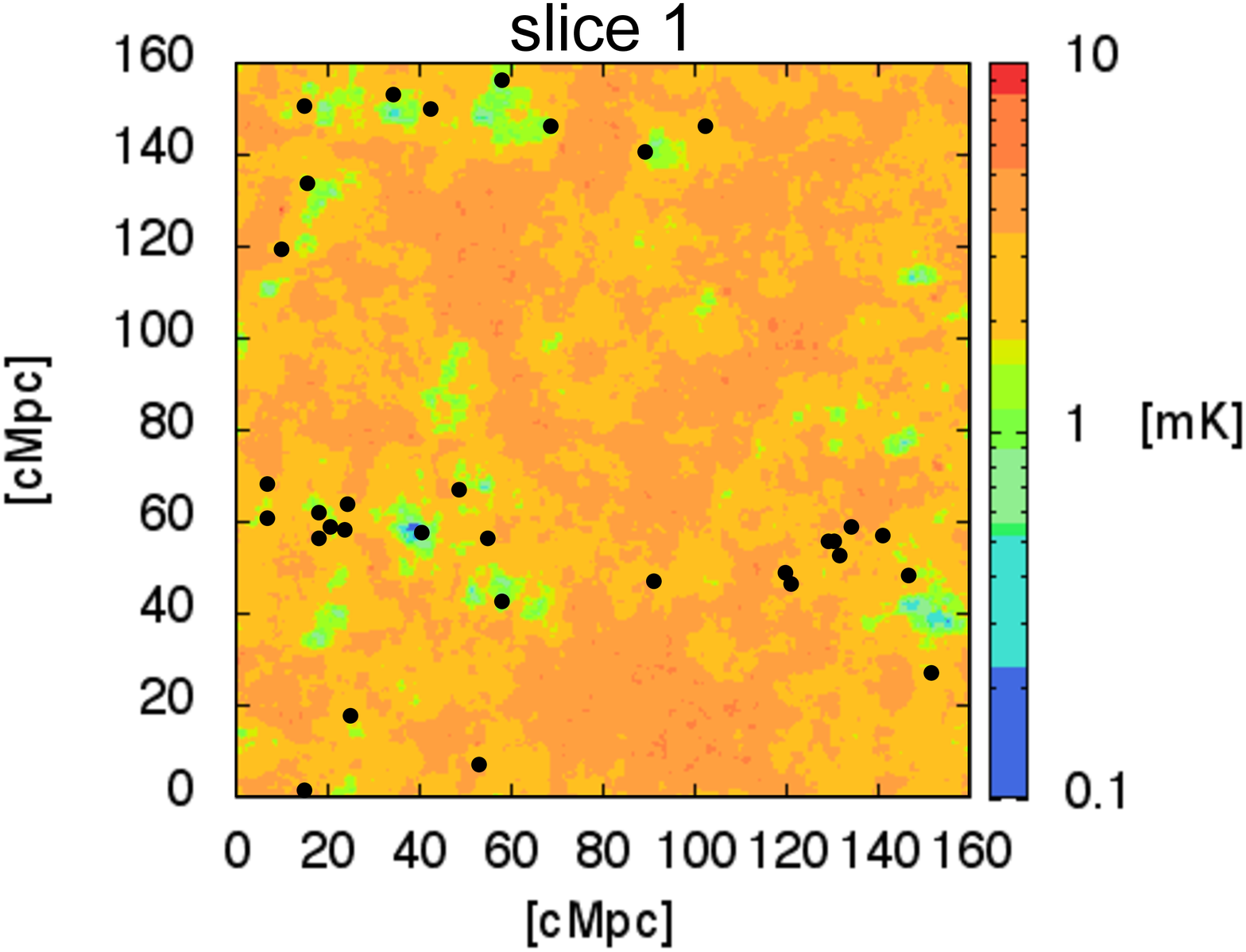}
        \end{center}
      \end{minipage}
      \begin{minipage}{0.35\hsize}
        \begin{center}
         \includegraphics[width=6.5cm]{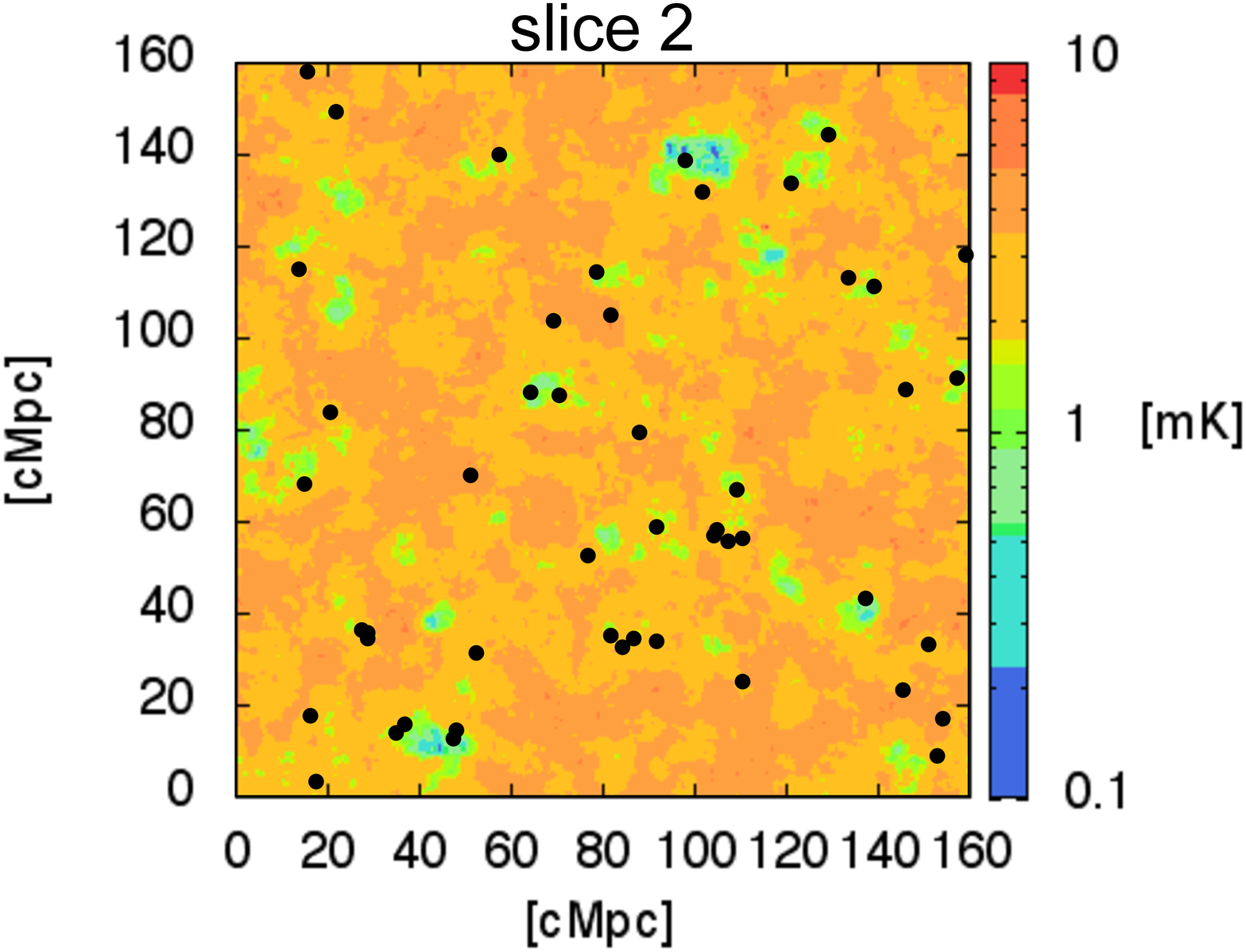}
        \end{center}
      \end{minipage}
\\      
      \begin{minipage}{0.35\hsize}
        \begin{center}
          \includegraphics[width=6.5cm]{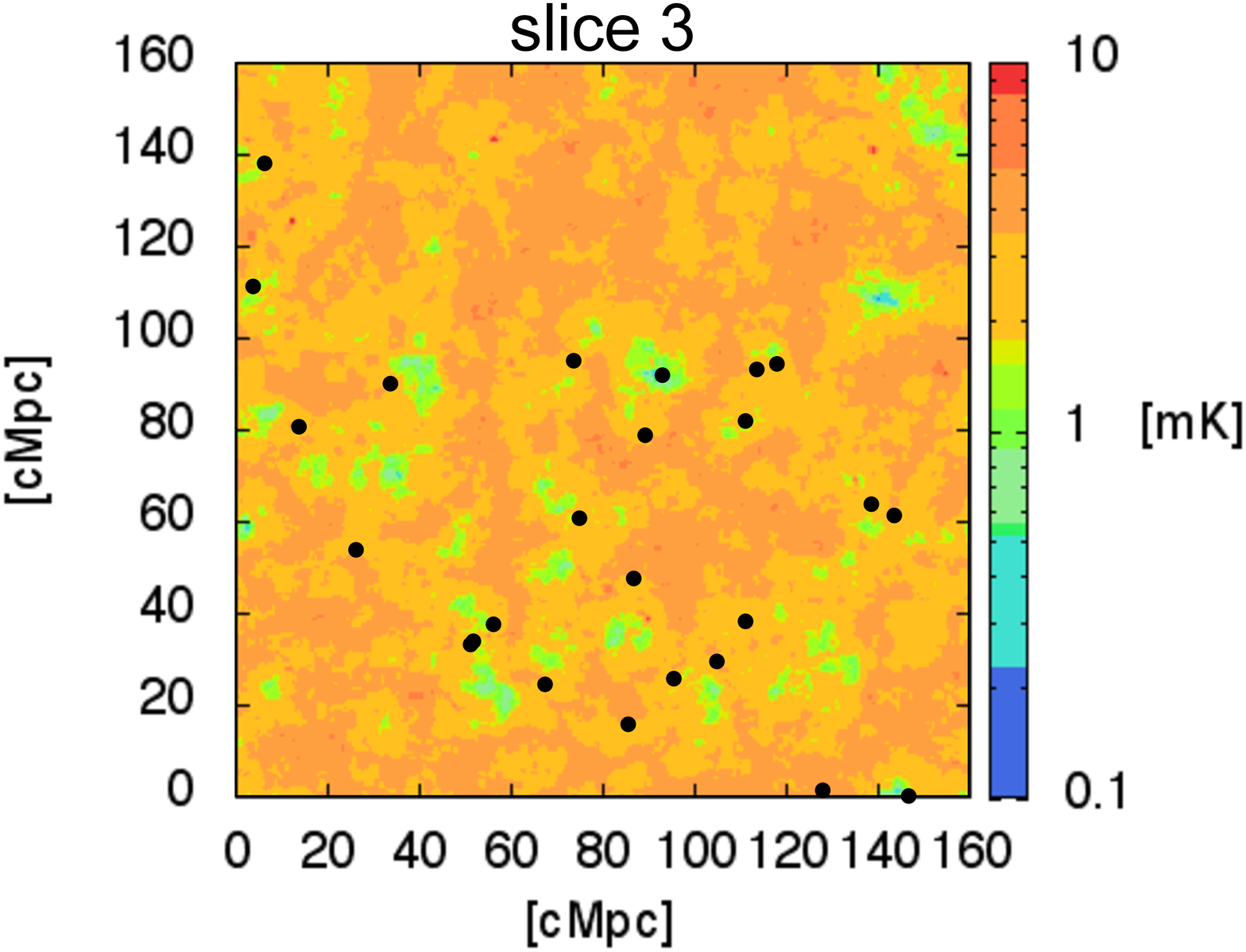}
        \end{center}
      \end{minipage}
      \begin{minipage}{0.35\hsize}
        \begin{center}
          \includegraphics[width=6.5cm]{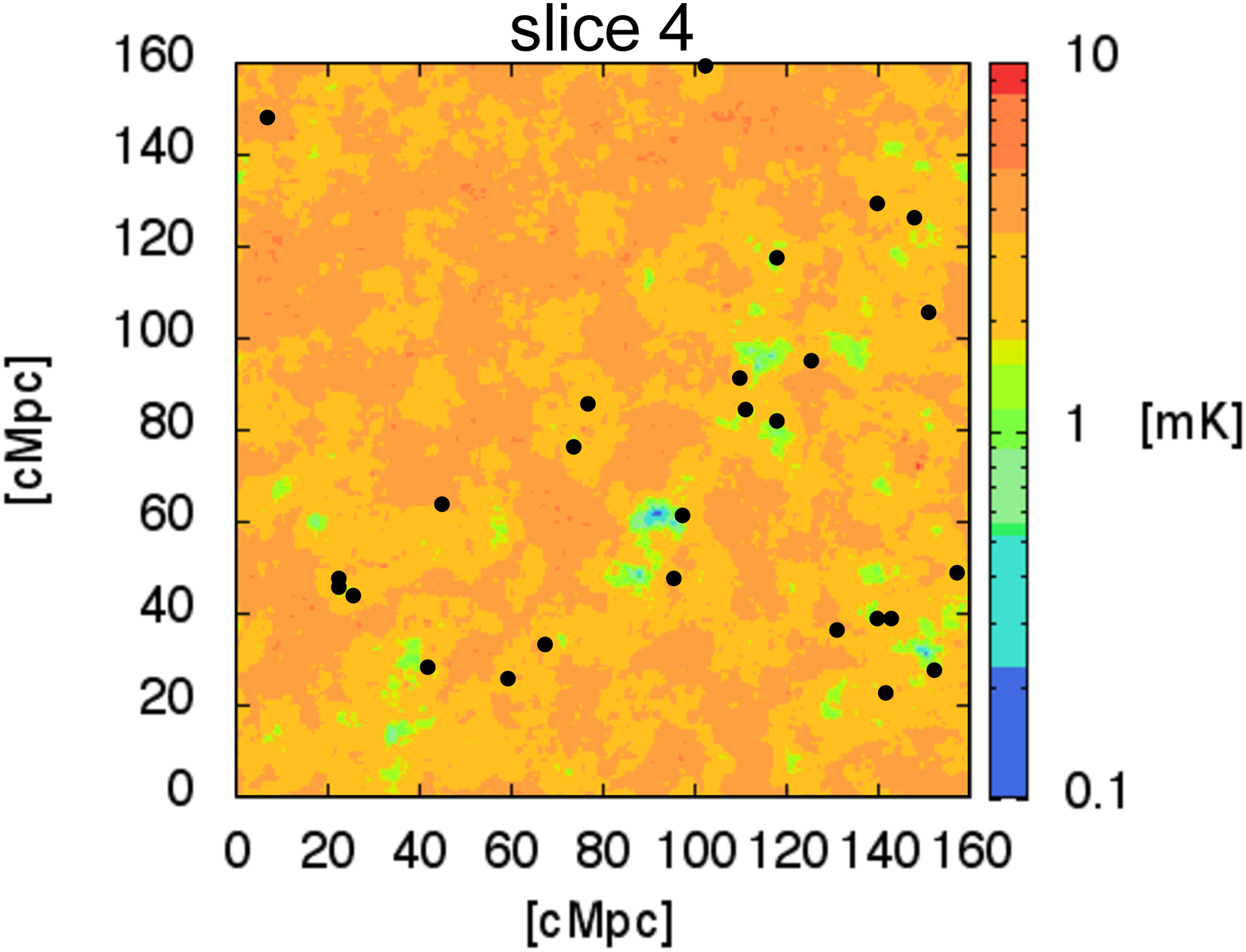}
        \end{center}
      \end{minipage}
   \end{tabular}
\caption{Same as Fig.~\ref{fig:mid_zp}, but for the `late' model.}
\label{fig:late_zp}
\end{center}
\end{figure*}

Similarly, Fig.~\ref{fig:mid_zp} and \ref{fig:late_zp} show LAE distributions for the spectroscopic LAE samples in 4 slices of the simulation box in the direction of z-axis.
Here, z is one of the 3 dimensions of the box and is different from the redshift.
In fact, the 4 slices have exactly the same redshift.
The width corresponds to the redshift uncertainty of the HSC survey $\sim 0.1$.
Exactly speaking, the HSC survey is performed along the light cone and there is a slight redshift evolution within a finite width of the light cone.
By using these slices of the simulation box, this redshift evolution is neglected.
As can be seen, the spectroscopic samples always produce smaller numbers of LAEs than the photometric samples.
We find number fractions of the spectroscopic LAEs to the photometric LAEs are $\sim 80\%$ in the LAE models of Model G.

\section{Error estimation}

\ \ The method for estimating a full error on the cross-power spectrum is the same manner as \citet{2007ApJ...660.1030F,2009ApJ...690..252L,2018MNRAS.479.2754K} and the details are described in \citet{2018MNRAS.479.2754K}.
The full error on the cross-power spectrum $\sigma_{\rm CPS}$ is determined by five cross-terms:
\begin{equation}
\sigma_{\rm CPS}\propto \sqrt{P^2_{21,\rm LAE}+P_{21}P_{\rm LAE}+P_{21}\sigma_{\rm g}+\sigma_{\rm N}P_{\rm LAE}+\sigma_{\rm N}\sigma_{\rm g}},
\end{equation}
where $P_{21}$ and $P_{\rm LAE}$ are 21cm-line and LAE auto power spectrum, respectively, and $\sigma_{\rm N}$ and $\sigma_{\rm g}$ are thermal noise on the 21cm-line observations and shot noise on the LAE survey, respectively.
Below, we regard the last term $\sigma_{\rm N}\sigma_{\rm g}$ as pure observational error, and the remaining four terms as the sample variance.

Concerning the HSC LAE survey, we consider Deep survey \citep{2018PASJ...70S..13O} where the total survey area is $27\ \rm deg^2$ and the minimum detectable Ly$\alpha$ luminosity is $4.1 \times 10^{42}\ \rm erg\ s^{-1}$.
For the photometric LAE sample, we estimate the error by assuming a redshift uncertainty of $\Delta z=0.1$, corresponding to the wavelength widths of the narrowband filters of HSC.
In the spectroscopic sample, we assume $\Delta z=0.0007$ \citep{2014PASJ...66R...1T} considering follow-up observations by the PFS.

To estimate the thermal noise of 21cm-line observation, $\sigma_{\rm N}$, we consider the MWA and SKA and assume that the MWA (SKA) has 256 (670) antenna tiles within $750\ (1,000)\ \rm m$, the effective area $14\ (462)\ \rm m^2$ per tile at $z=6.6$, the bandpass $8\ (8)\ \rm MHz$ and field-of-view $800\ (25)\ {\rm deg^2}$. 
It is assumed that both telescopes observe the HSC field for 1,000 hours with the central redshift of 6.6.
The field area of cross-correlation analysis is the smaller of the HSC and 21cm-line fields, which is $27\ {\rm deg^2}$ and $25\ {\rm deg^2}$ for MWA+HSC and SKA+HSC, respectively.

\section{Results}

\subsection{Cross-power spectrum signal in Model G}

First of all, we compare the 21cm-LAE cross-power spectrum signal of photometric LAE samples between Paper I model and Model G of \citet{2018PASJ...70...55I} \footnote{In Paper I, we assumed that all the photometric LAEs can be spectroscopically confirmed by the PFS.
However, as we mentioned before, only $80\%$ of the photometric samples are included in the spectroscopic samples.
To make a fair comparison, we demonstrate only 2D cross-correlation spectra using the photometric samples here.}.
In the photometric samples, only 2D cross-power spectrum can be measured since the precise redshifts of the LAEs are not available.
To estimate the 2D cross-power spectrum, we integrate the 21cm line signal and LAEs within $\Delta z=0.1$, which corresponds to the redshift uncertainty of HSC, along the redshift direction.

\begin{figure}
\begin{center}
\includegraphics[width=8cm]{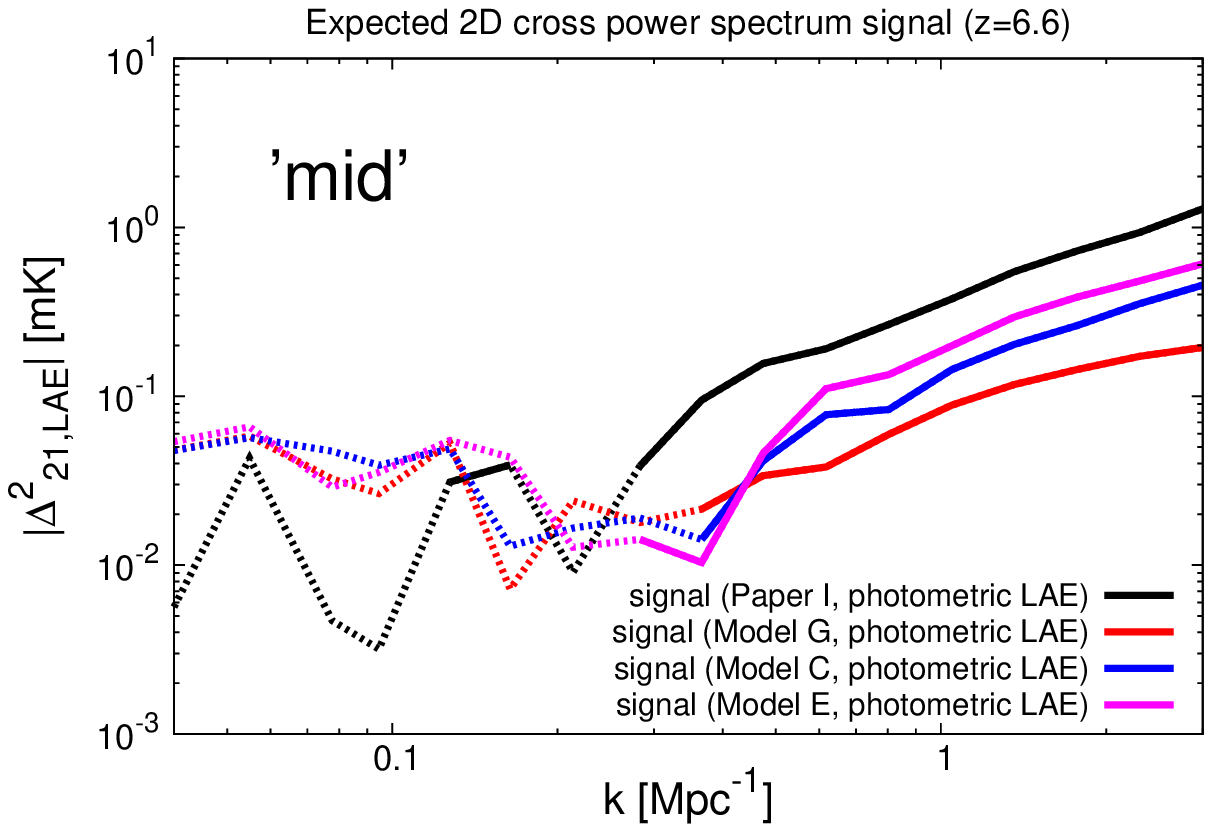}
\includegraphics[width=8cm]{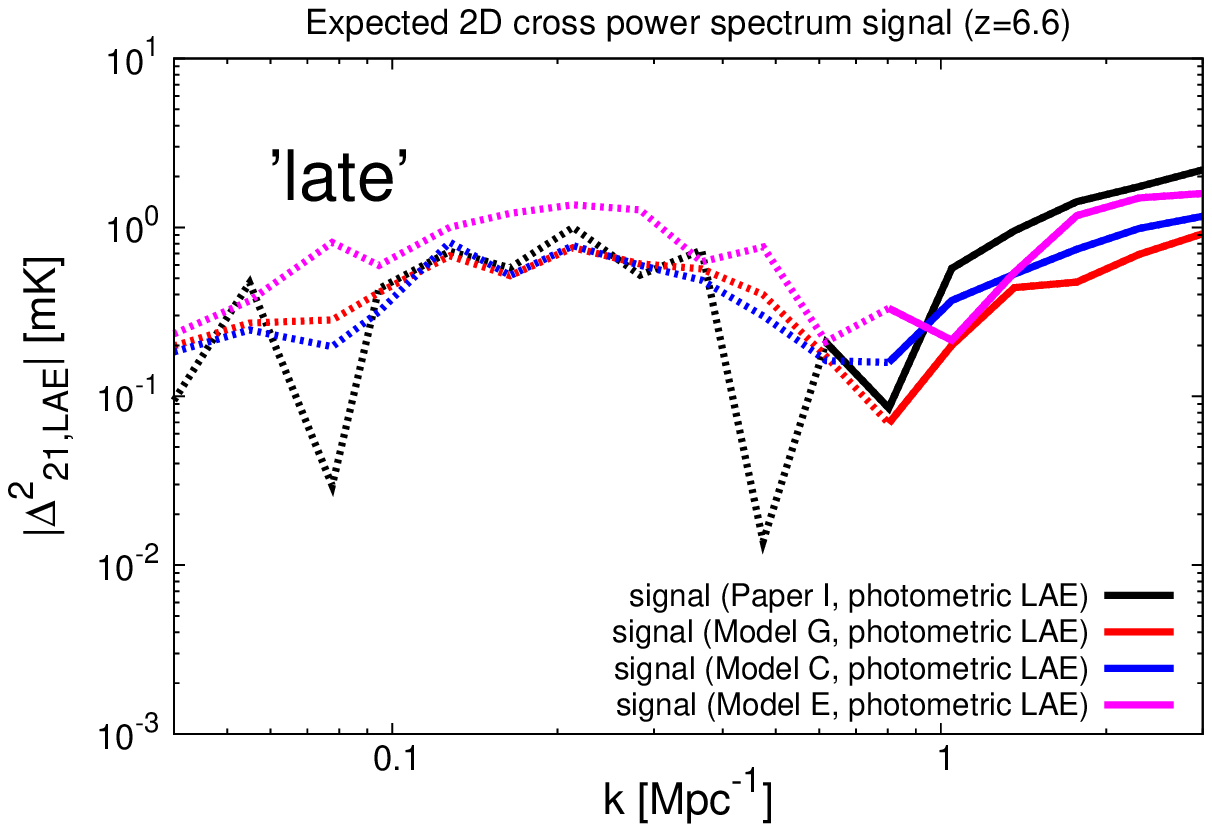}
\end{center}
\caption{Comparison of the averaged 2D 21cm-LAE cross-power spectrum with the LAE model in Paper I (black), Model G (red), Model C (blue) and Model E (magenta) at $z=6.6$, where LAEs are photometrically identified. The solid lines represent positive values and the dotted lines represent negative values. (Top) `mid' model. (Bottom) `late' model.}
\label{fig:signal_mag}
\end{figure}

Fig.~\ref{fig:signal_mag} shows the 21cm-LAE cross-power spectra of Paper I model and Model G at $z=6.6$ in the cases of `mid' and `late'.
The cross-power spectra with Model G are well consistent with those adopting Paper I model at large scales ($k\lesssim0.2\ \rm Mpc^{-1}$ in the `mid' model and $k\lesssim0.4\ \rm Mpc^{-1}$ in the `late' model).
This implies the cross-correlation signal is not so sensitive to the LAE models at the large scales.
The stochasticity and halo mass dependence, which are accounted in Model G, could change the clustering feature of the LAEs at small scales, but the effect is not significant for large-scale fluctuations.
The clustering feature affects the cross-correlation power at smaller scales than the turnover scale where the power is positive due to the correlation between the galaxies and the density field.
Contrastingly, the large-scale power is sensitive to the reionization model rather than the LAE model.
Thus, the two models converge at larger scales than the turnover scale.
This trend can be also found in \citet{2016MNRAS.459.2741S}.
Indeed, the consistency of the cross-correlation signals among the LAE models implies a simple LAE model is still acceptable for the prediction of large-scale cross-correlations.

The biggest difference of the cross-power spectra between the LAE models is a signal loss at small scales.
In the `mid' model, the amplitude of the cross-power spectrum adopting Model G is smaller than that adopting the simple LAE model by one order of magnitude at $k\sim1\ \rm Mpc^{-1}$.
We can see the signal loss by a few factors of magnitude at $k \sim 1\ \rm Mpc^{-1}$ in the `late' model as well.
This indicates adopting an appropriate LAE model is important to predict the small-scale cross-power spectrum, and the signal loss could affect the study of the detectability.

In Fig.~\ref{fig:signal_mag}, the cross power spectra of Models C and E are also shown.
They have the similar behavior and amplitude to those of Model G at large scales for both `mid' and `late' models.
On the other hand, at small scales of $k \gtrsim 1\ \rm Mpc^{-1}$, Models C and E have smaller power than Paper I model but larger power than Model G.
The power of Model E is slightly larger than that of Model C and the difference is less than 30\%. This indicates that both the stochasticity and halo-mass dependence of Ly$\alpha$ escape fraction comparably contribute to the suppression of the cross power spectrum of Model G at small scales.

\begin{figure}
\begin{center}
\includegraphics[width=10cm]{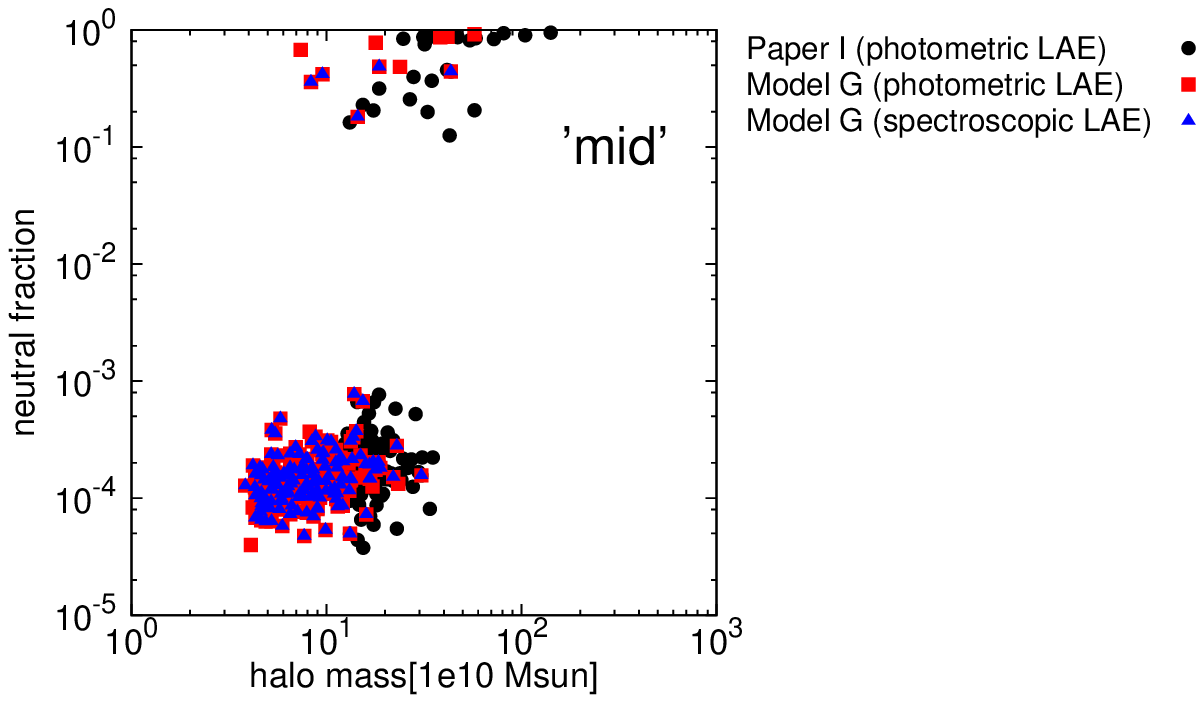}
\includegraphics[width=10cm]{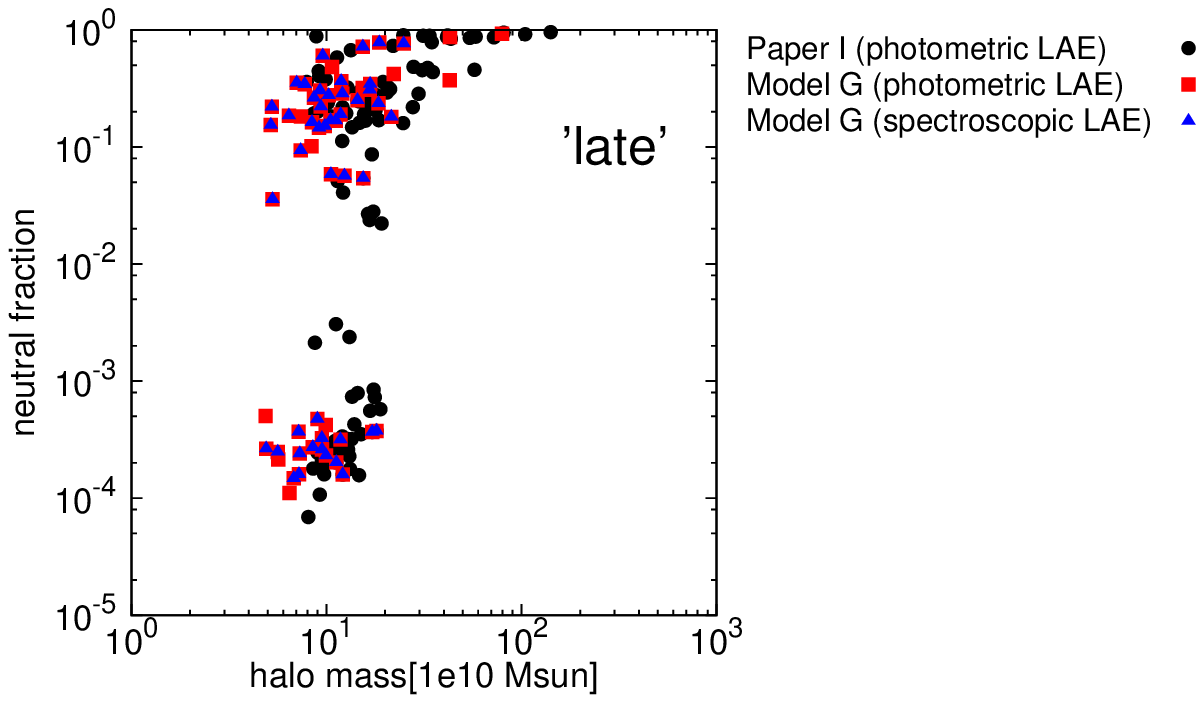}
\end{center}
\caption{The halo mass of LAEs and the neutral fraction in the simulation grid where the LAEs reside. The black circle symbols represent the photometric LAEs in Paper I model. The red square and blue triangle symbols represent the photometric and spectroscopic LAEs in ModelG, respectively. The realization of ModelG set1 is shown, as an example. (Top) `mid' model. (Bottom) `late' model.}
\label{fig:LAE_halo}
\end{figure}

To explain the difference in the cross-power spectrum at the small scales, we show a scatter plot of the halo mass and neutral fraction of the grids hosting LAEs in Fig.~\ref{fig:LAE_halo}.
Comparing Paper I with Model G, it is seen that the LAEs identified in Paper I are galaxies in massive halos surrounded by the IGM with a high neutral fraction\footnote{The reason why our simulation predicts a higher neutral hydrogen fraction around massive halos is probably that we assumed an ionizing photon escape fraction to be small in massive halos. We note the scatter plot could depend on not only LAE models but also reionization simulations.}.
Such galaxies are located at high density regions and the strong clustering enhances the cross-power spectrum at the small scales.
In Model G, Ly$\alpha$ photons from such massive galaxies have higher optical depth because of the stochasticity and the halo-mass dependence (see Eqs. (\ref{p_tau}) and (\ref{tau_halomass})).
Thus, such galaxies are not identified as LAEs in Model G.

\subsection{Detectability for Model G}

Here, We discuss the detectability of cross-power spectrum with the photometric and spectroscopic LAE samples in Model G.

\subsubsection{Photometric LAE samples}

Fig.~\ref{fig:error_mag_shade} shows the cross-power spectra, the full errors, and the observational errors ($\sigma_{\rm N}\sigma_{\rm g}$) for MWA and SKA in the `mid' and `late' models, respectively.
In the `mid' model, the negative correlation of 2D cross-power spectrum could be detectable at large scales ($k \lesssim 0.2\ \rm Mpc^{-1}$) since the full error is enough small, while the signal is comparable to the observational error at around the turn over scale.
Unfortunately, the combination of MWA and HSC has a severe difficulty to detect the signal.

In the `late' model, thanks to the large amplitude of the signal, the MWA and HSC cross-correlation has the sensitivity comparable to the signal, but the large sample variance makes the detection less likely.
On the other hand, the detectability of SKA extends to smaller scales ($k\lesssim0.5\ \rm Mpc^{-1}$).
However, because of the large redshift uncertainty of HSC, the observational errors are much larger than the signal at small scales ($k\geq0.8\ \rm Mpc^{-1}$).
Consequently, the small-scale signature shown in Sec.5.1 would not be observable even with the combination of SKA and HSC.

\subsubsection{Spectroscopic LAE samples}

Next, we discuss the cases with spectroscopic LAE samples.
In this case, 3D cross-power spectrum can be measured since the precise redshifts of the LAEs are available.
Here, to reduce statistical uncertainty in the estimation of 3D cross-power spectrum, we generate 12 data sets from our simulation box by dividing the box into 4 slices with respect to $x, y, z$-axes, respectively, and we take the average value of the signals.
The each slice is equivalent to a single survey volume at redshift $z=6.6$ with a width of $40~{\rm Mpc}$, corresponding to the HSC redshift uncertainty.
Note that, although the 12 data sets are generated from a single realization of our simulation and, therefore, not fully statistically independent, to treat them as independent samples is reasonable at scales smaller than the slice width ($k \lesssim 0.1~{\rm Mpc}^{-1}$.

\begin{figure}
\begin{center}
\includegraphics[width=8cm]{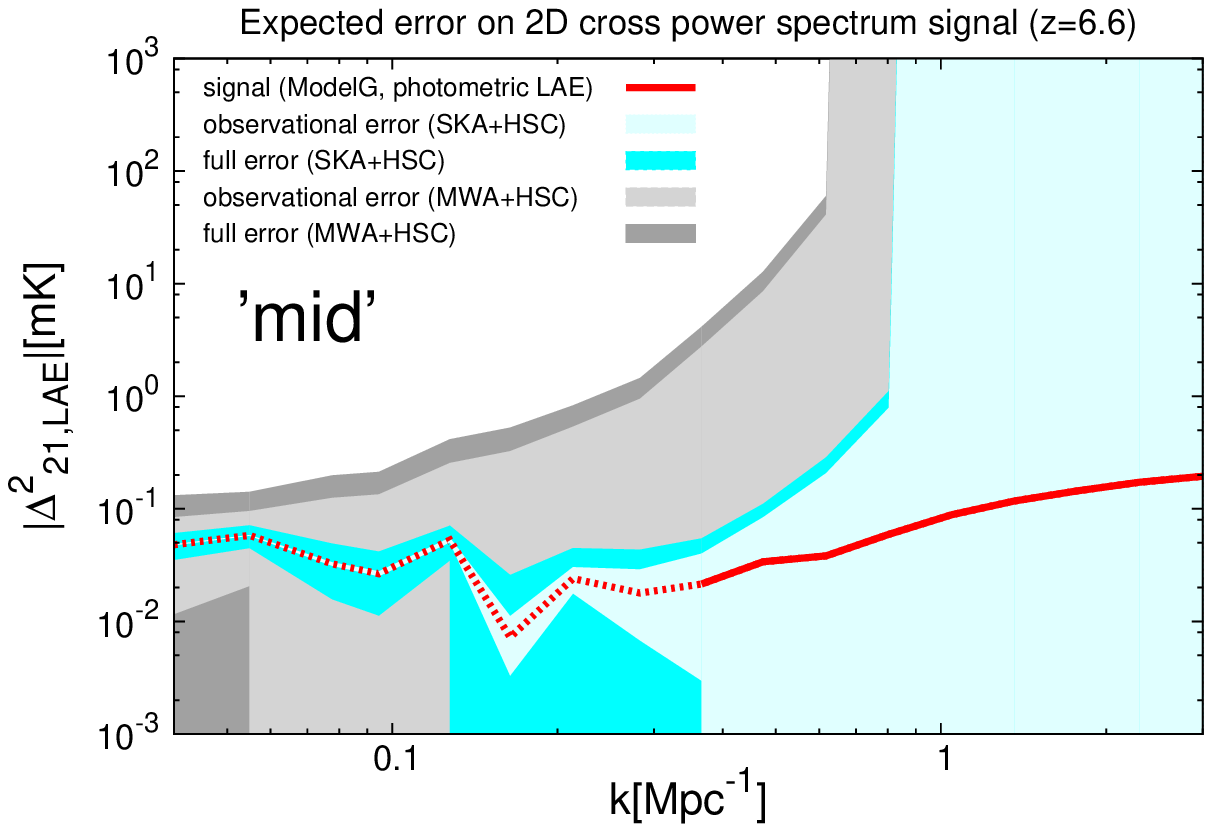}
\includegraphics[width=8cm]{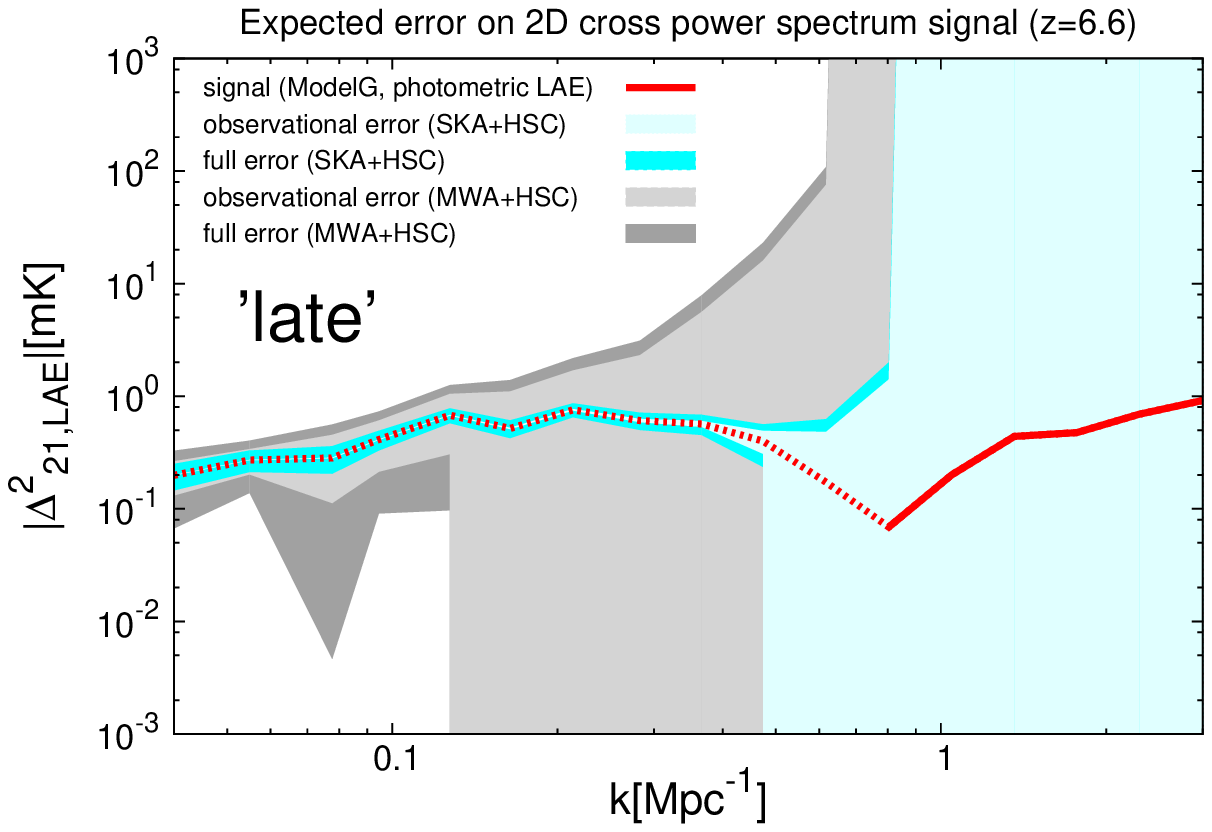}
\end{center}
\caption{The averaged 2D 21cm-LAE cross-power spectrum and the expected error at $z=6.6$, where LAEs are photometrically identified in Model G. The red curve shows the cross-power spectrum signal and the light-gray (-cyan) shading shows the observational error for the MWA (SKA). The dark-gray (-cyan) shading shows full error including the sample variance for the MWA (SKA)$\times$HSC Deep survey. (Top) `mid' model. (Bottom) `late' model.}
\label{fig:error_mag_shade}
\end{figure}

\begin{figure}
\begin{center}
\includegraphics[width=8cm]{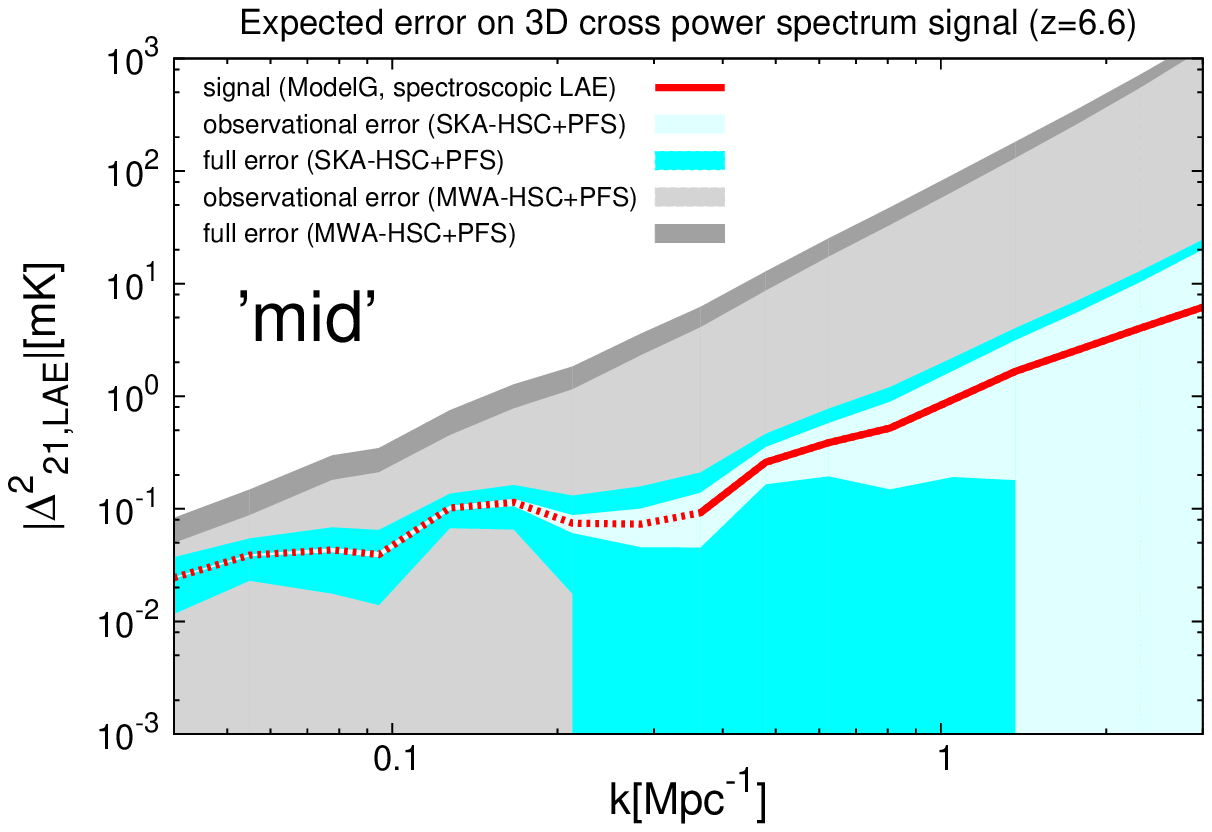}
\includegraphics[width=8cm]{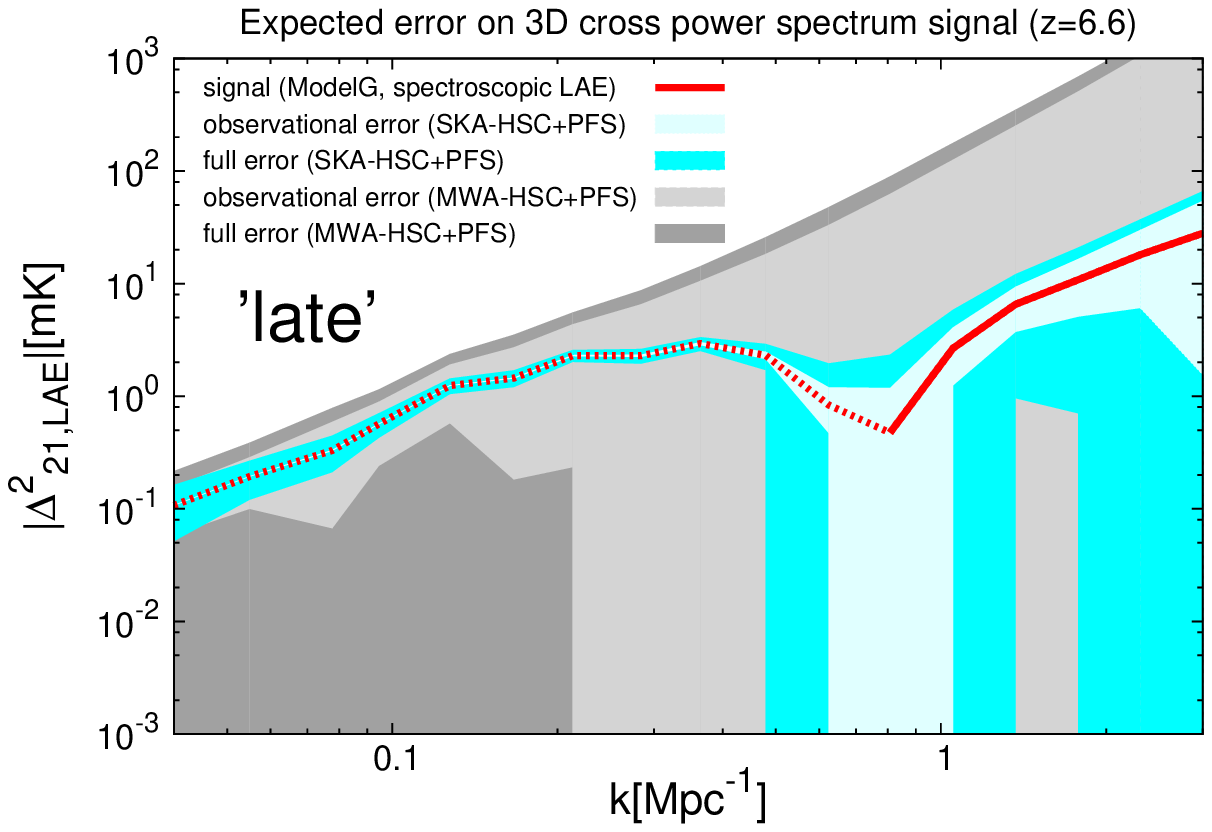}
\end{center}
\caption{Same as Fig.~\ref{fig:error_mag_shade}, but LAEs are spectroscopically identified in Model G.}
\label{fig:error_dz_EW_shade}
\end{figure}

Fig.~\ref{fig:error_dz_EW_shade} shows the 3D cross-power spectra, the full errors, and the observational errors ($\sigma_{\rm N}\sigma_{\rm g}$) for MWA and SKA in the `mid' and `late' models, respectively.
The amplitudes at large scales are consistent with the 2D signal in Fig.~\ref{fig:error_mag_shade}, while the signal is enhanced at small scales for the 3D case.
This behavior is natural because small-scale fluctuations are washed out if the redshift error is as large as given by the HSC only.
Thus, the detectability of the signal does not change at large scales.
In addition, thanks to the small redshift uncertainty of the PFS, the observational errors at small scales ($k \geq 0.8\ \rm Mpc^{-1}$) are drastically improved in the both cases of the MWA and SKA.
However, a statistically significant detection at these scales is still hard even for the SKA.

\subsubsection{Requirement for detection of the small scale signature}

As we have seen, the signal amplitude at small scales of our best model is suppressed compared with that of Paper I model and the expected signal-to-noise ratio is reduced accordingly.
The detection of small-scale signal is important to detemine the turnover scale and probe a typical size of ionized bubbles.
In Paper I, we reported that an effective way to enhance the detectability is to expand the survey area rather than to perform a deeper observation to detect fainter LAEs for a fixed total observation time.
Here, we investigate the effect of expanded survey area and deeper 21cm-line observation to enhance the detectability at small scales taking the `mid' model as an example.

\begin{figure}
\begin{center}
\includegraphics[width=8cm]{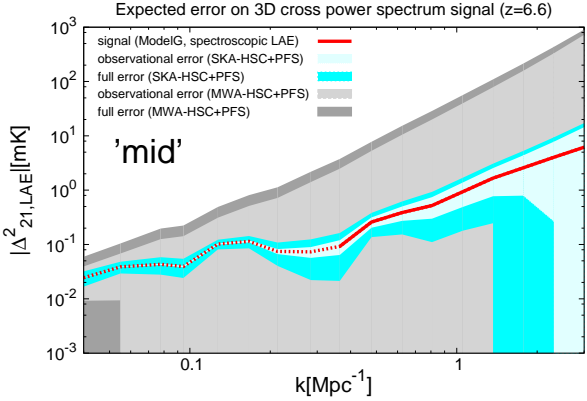}
\end{center}
\caption{Same as Fig.~\ref{fig:error_dz_EW_shade} in the `mid' model, but with an extended LAE survey area to $75\ \rm deg^2$.}
\label{fig:error_dz_EW_75deg}
\end{figure}

Fig.~\ref{fig:error_dz_EW_75deg} shows the 3D cross-power spectrum, the sample variance and the observational errors in the `mid' model, where the survey area is expanded to $75\ \rm deg^2$.
Here, the survey time per pointing (survey depth) is fixed so that the total survey time increases by a factor of 3 for the HSC and SKA, while the observation time for the MWA is unchanged because it has even larger field-of-view ($800\ \rm deg^2$).
In this case, the observational error in the SKA and HSC cross-correlation is smaller than the cross-power spectrum signal, and the sample variance is enough small to identify the signal up to $k \lesssim 1\ \rm Mpc^{-1}$.
Therefore, the increase in the LAE survey area by a factor of 3 is enough to determine the turnover scale.

\begin{figure}
\begin{center}
\includegraphics[width=8cm]{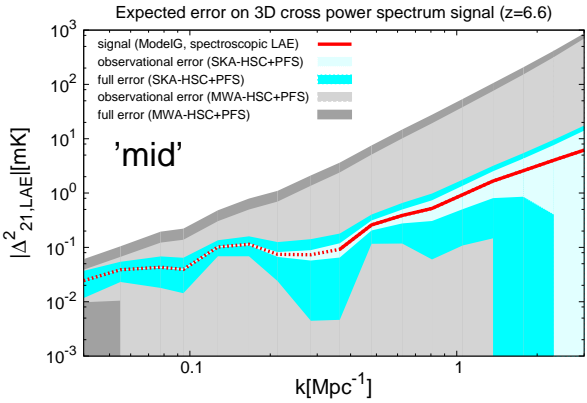}
\end{center}
\caption{Same as Fig.~\ref{fig:error_dz_EW_shade} in the `mid' model, but the 21cm observations are extended to 3000 hrs observation time.}
\label{fig:error_dz_EW_3000hrs}
\end{figure}

Next, Fig.~\ref{fig:error_dz_EW_3000hrs} shows the 3D cross-power spectrum, the full observational errors and the observational errors, where the 21cm-line observation time is extended to 3,000 hrs.
This case also enables the SKA and HSC to detect the positive correlation signal at $k \sim 1\ \rm Mpc^{-1}$. Thus, extended observation time of either the HSC or the SKA is enough to detect small-scale signal with a high statistical significance and probe the turnover scale.

\section{Summary and Discussion}

In this paper, we have revisited the detectability of the 21cm-LAE cross-power spectrum adopting a state-of-the-art model of LAE distribution developed by \citet{2018PASJ...70...55I} that is consistent with all Subaru/HSC observations such as the Ly$\alpha$ luminosity function, LAE angular auto-correlation and the LAE fractions in the continuum selected galaxies.
We presented the 21cm-LAE cross-power spectrum signals and compared with our previous model. Then we estimated the observational errors for the photometric LAE samples and spectroscopic LAE samples.
As a result, we found the cross-power spectrum at the small scales ($k \sim1 \ \rm Mpc^{-1}$) is sensitive to the details of LAE models, and the amplitude is smaller for the updated LAE models.
One of our conclusion is that appropriate LAE models are required for the precise prediction of cross-power spectra at the small scales.
Further, we found that the small-scale signals are hard to detect even with the SKA and HSC even if PFS redshifts are available, but an extended HSC survey with a larger survey area by a factor of 3 or an extended SKA with 3000 hrs observation time will be enough to measure cross-power spectra at as small scales as $k \sim 1~{\rm Mpc}^{-1}$.
On the other hand, the cross-power spectrum at the large scales is less sensitive to the details of LAE models.
Thus, simple LAE models considered so far are enough to predict the expected signal at large scales ($k\lesssim0.2\ \rm Mpc^{-1}$), and the discussion of the detectability at the large scales in Paper I is valid.

Finally, we discuss the feedback on the LAE models from the cross-correlation observations.
As we saw above, we found the difference of the LAE models appears in the amplitude of the cross-power spectrum at small scales.
In the updated LAE model (Model G) with the stochasticity and halo mass dependence of Ly$\alpha$ escape fraction, the typical halo mass hosting LAEs is smaller, which results in the lower amplitude of the cross-correlation signal at the small scales.
In fact, the correlation signal loss at the small scales has been found in the LAE angular auto-correlation.
\citet{2018PASJ...70...55I} has shown that Model A, which is essentially the same model as Paper I model, predicts a larger amplitude in smaller angular separations ($< 60\ \rm arcsec$) than Model G.
Thus, the typical halo mass may truly be suppressed by the stochasticity and halo mass dependence of Ly$\alpha$ escape fraction.
However, there is still a possibility that the observed LAEs are hosted by massive halos because the measured LAE angular auto-correlation has a large uncertainty.
Therefore, if the small-scale cross-correlation is measured, we can probe the typical halo mass of LAEs and confirm the implication from the measurement of the LAE auto-correlation.

\section*{Acknowledgement}
We would appreciate the constructive comments from anonymous referee which were very useful to improve the quality of this paper.
We also thank Tomoaki Ishiyama for providing us with $N$-body simulation data used in this work, and thank Shintaro Yoshiura for giving us some useful comments.
Numerical simulations were carried out on CrayXC30 installed at Center for Computational Astrophysics of National Astronomical Observatory of Japan, NAOJ.
This work is supported by Grand-in-Aid from the Ministry of Education, Culture, Sports, and Science and Technology (MEXT) of Japan, No.26610048, No.15H05896, No.16H05999, No.17H01110 (KT, KH), No.17H01114 (AKI), JP18K03699 (KH), Bilateral Joint Research Projects of JSPS (KT), and a grant from NAOJ.




%
%


\bsp	
\label{lastpage}
\end{document}